\documentclass{emulateapj}

\usepackage{amssymb}
\usepackage{amsmath}
\usepackage{natbib}
\usepackage{txfonts}
\usepackage[colorlinks=true,citecolor=blue,linkcolor=blue]{hyperref}
\usepackage{microtype}
\usepackage{color}
\usepackage{graphicx}
\usepackage{epstopdf}

\def\apjl{ApJL}
\def\apjs{ApJS}

\def\aap{A\&A}

\usepackage{graphicx}
\usepackage{dcolumn}
\usepackage{bm}
\usepackage{color}
\usepackage{soul}
\renewcommand{\thefootnote}{\fnsymbol{footnote}}
\usepackage{mathtools}
\usepackage[T1]{fontenc}

\def\camk{1}
\def\soton{2}

\shorttitle{Tidal deformations of hybrid stars with sharp phase transitions and elastic crusts}
\shortauthors{Pereira, Bejger, Andersson and Gittins}

\begin{document}

\title{Tidal deformations of hybrid stars with sharp phase transitions and elastic crusts}

\author{Jonas P.~Pereira\altaffilmark{\camk}, Micha{\l} Bejger\altaffilmark{\camk}, Nils Andersson\altaffilmark{\soton}, and Fabian Gittins\altaffilmark{\soton}}

\altaffiltext{\camk}{Nicolaus Copernicus Astronomical Center, Polish Academy of Sciences, Bartycka 18, 00-716, Warsaw, Poland}
\altaffiltext{\soton}{Mathematical Sciences and STAG Research Centre,
University of Southampton, Southampton, SO17 1BJ, United Kingdom}

\begin{abstract}
Gravitational wave astronomy is expected to provide independent constraints on neutron star properties, such as their equation of state. This is possible with the measurements of binary components' tidal deformability, which alter the point-particle gravitational waveforms of neutron-star binaries. Here we provide a first study of the tidal deformability effects due to the elasticity/solidity of the crust (hadronic phase) in a hybrid neutron star, as well as the influence of a quark-hadronic phase density jump on tidal deformations.
We employ the framework of nonradial perturbations with zero frequency and study hadronic phases presenting elastic aspects when perturbed (with the shear modulus approximately $1\%$ of the pressure).
We find that the relative tidal deformation change in a hybrid star with a perfect-fluid quark phase and a hadronic phase presenting an elastic part is never larger than about $2-4\%$ (with respect to a perfect-fluid counterpart). These maximum changes occur when the elastic region of a hybrid star is larger than approximately $60\%$ of the star's radius, which may happen when its quark phase is small and the density jump is large enough, or even when a hybrid star has an elastic mixed phase. For other cases, tidal deformation changes due to an elastic crust are negligible ($10^{-5}-10^{-1}\%$), therefore unlikely to be measured even with third generation detectors. Thus, only when the size of the elastic hadronic region of a hybrid star is over half of its radius, the effects of elasticity could have a noticeable impact on tidal deformations.
\end{abstract}


\altaffiltext{}{jpereira@camk.edu.pl}

\maketitle

\section{Introduction}

The recent detection of gravitational waves (GWs) from binary compact systems \citep{2016PhRvL.116f1102A,2016PhRvL.116x1103A,PhysRevLett.118.221101,2017PhRvL.119p1101A,2018arXiv180511581T,2019PhRvX...9c1040A,2020arXiv200101761T} has inaugurated the era of GW astronomy, and  we now have the real possibility of addressing many issues for compact astrophysical objects. After the first detection of GWs from neutron stars (NSs) \citep{2017PhRvL.119p1101A}, independent constraints from X-ray observations---which rely on pulse profile modelling (see, e.g., \citealt{2016ApJ...832...92O,Miller_2019,Riley_2019,2020ApJ...889..165D} and references therein)---have already been obtained, for instance on their equation of state (EOS) and radii \citep{2018PhRvL.120q2703A,2018PhRvL.120z1103M,PhysRevLett.121.091102,2018ApJ...857...12N,Paschalidis2018}. Further constraints are possible due to inferences of tidal deformations \citep{2008ApJ...677.1216H,2009PhRvD..80h4035D,2010PhRvD..81l3016H}, which already have known upper limits \citep{2017PhRvL.119p1101A,2018arXiv180511581T,2019PhRvX...9a1001A}. Future GW detections hold the promise of stronger constraints on the NS parameters \citep{2017CQGra..34d4001A,2017ApJ...837...67C,2019PrPNP.10903714B}. We focus here on whether GW observations may be able to provide evidence for elastic and hybrid phases in NS interiors.

Hybrid NSs are compact systems presenting both quark and hadronic phases. They might be possible in nature due to the existence of phase transitions in quantum chromodynamics (QCD, see \citealt{2008RvMP...80.1455A,Paschalidis2018,2018arXiv180901116B,Alford:2019oge}). However, the order of this phase transition is unknown in the parameter range relevant for NSs (low temperatures and large chemical potentials) \citep{2008RvMP...80.1455A}. In addition, since the surface tension for dense matter is poorly known, the possibility of a mixed phase remains an open issue \citep{Bombaci:2016xuj,Lugones:2016ytl}. Astrophysical observations cannot currently exclude hybrid NSs \citep{2017A&A...600A..39B}. However, this might be possible with future, more precise GW measurements from the late inspiral and also the postmerger phase \citep{2018ApJ...857...12N,2018arXiv180901116B,2018PhRvL.120z1103M}. Another promising way of probing hybrid stars is through pulse profile modelling, timing and polarimetry measurements i.e., with the NICER \citep{2016ApJ...832...92O} and eXTP \citep{2016SPIE.9905E..1QZ} missions. The former has already provided the first measurements of an NS radius with an uncertainty of around $10\%$ at the 68\% confidence level \citep[see][]{Bilous_2019,Riley_2019,Raaijmakers_2019,Miller_2019,Bogdanov_2019,Bogdanov_2019_,Guillot_2019}. The eXTP mission promises even more constraining measurements (${\sim}5\%$ uncertainties or less) \citep{2016SPIE.9905E..1QZ}. NICER's precision is already in an interesting range to probe the interiors of NS (see e.g. \citealt{2018A&A...616A.105S}), especially hybrid stars that allow for additional stable branchs (``third family'') with smaller radii than their hadronic counterparts \citep{2017PhRvC..96d5809A,2017PhRvL.119p1104A,2019arXiv191209809C}.

It is already known that tidal deformations of hadronic stars with solid/elastic crusts and liquid cores hardly differ at all from perfect-fluid stars in ordinary cases \citep{2011PhRvD..84j3006P,2020arXiv200305449G},  which assume a continuous energy density at the interface between a (sizable) core and a solid crust. However, the situation may be different for hybrid stars. As is well known, the presence of density jumps in hybrid stars with sharp phase transitions strongly influences  the tidal deformation in the case of perfect fluids (see, e.g., \citealt{2009PhRvD..80h4035D,2010PhRvD..82b4016P,2019PhRvD..99h3014H}). To the best of our knowledge, studies have not been carried out probing the consequences of
combined density jumps and shear stresses in the hadronic phase of a hybrid NS. Therefore, we investigate how this issue impacts on GW observations. The question is complementary to the existing studies of hybrid stars in the context of GW170817 \citep{2019arXiv191100276J,2019PhRvD..99j3009M,2019arXiv191009740E,Alford:2019oge,2019PhRvD..99h3014H,2019PhRvL.122f1102B,Paschalidis2018,2019A&A...622A.174S}, which assume perfect-fluid systems. Natural motivations for our study are that a discontinuous quark-hadronic density profile might non-negligibly change the tidal deformability of solid/elastic hybrid stars and a nonzero shear modulus right from the bottom of the hadronic phase might be seen as a simplistic model for an elastic mixed phase.

It is also known that solid aspects of a quark phase could significantly change the tidal deformation of NSs \citep{2017PhRvD..95j1302L,2019PhRvD..99b3018L}. QCD allows for such a possibility through the so-called LOFF phase, where shear stresses could be up to a thousand times larger than crustal ones \citep{2007PhRvD..76g4026M}. It has been reported that for purely crystalline quark stars tidal deformations could decrease up to $60\%$ with respect to their perfect fluid counterparts \citep{2017PhRvD..95j1302L}. On the other hand, if the quark phase is located inside a hadronic phase, tidal deformations are much subtler: (i) if the (elastic) quark phase extends beyond around $70\%$ of the whole star, it still follows that the tidal deformations could change considerably; (ii) in other cases, changes with respect to a perfect fluid are expected to be small \citep{2019PhRvD..99b3018L}.

For the above case in particular, lower limits on the tidal deformations coming from GW170817 \citep{2018ApJ...852L..29R,2019MNRAS.489L..91C,2019EPJA...55...50R,2019ApJ...876L..31K}  are important for probing hybrid star models and the solidity of the quark phase. In addition, maximum deformations of stars with elastic quark phases might lead to observable effects in current GW detectors \citep{2007PhRvL..99w1101H}, and may allow for the estimation of breaking strains of solid stars \citep{2000MNRAS.319..902U}. Although the issue of the maximum deformation of solid stars is very important for determining their maximum ellipticities in general, in this work we only focus on the cases of small deformations of elastic hadronic regions of stars (solid/elastic crusts). The reason for doing so is that the elastic properties of such matter are currently better understood.

The article is arranged as follows: in Sec. \ref{formalism} we review the main ingredients for calculating tidal deformations of perturbed stars with static tidal deformations. Sec. \ref{model} is devoted to presenting simple quark and hadronic models for a hybrid star, which can be used to estimate tidal deformations when part of the hadronic phase is elastic. In Sec. \ref{boundary-conditions} we obtain the appropriate boundary conditions for hybrid stars presenting sharp phase transitions with no mixed phases. Our results regarding tidal deformations of hybrid stars with solid crusts are presented in Sec. \ref{results}. We discuss the main issues raised by our work, and provide the relevant conclusions, in Sec. \ref{discussion}.

We work with geometric units. Our metric signature convention is $+2$. Unless otherwise stated, $\Delta X$ is defined as the Lagrangian perturbation \citep{1986bhwd.book.....S,2007LRR....10....1A,2011PhRvD..84j3006P,2019CQGra..36j5004A} of a physical quantity $X$, while $\delta X$ is the corresponding Eulerian perturbation.

\section{Formalism}
\label{formalism}

The formalism we follow has been laid down by \citet{2011PhRvD..84j3006P}. Here we briefly review it, focusing on the equations we will use (for further details, see \citealt{2020arXiv200305449G}).
For the background hybrid stars, fluids are assumed to be perfect and hence described by the Tolman-Oppenheimer-Volkoff (TOV) system of equations, namely,
\begin{eqnarray}
\label{tov1}
&&\frac{dp}{dr} = - \frac{\epsilon m}{r^2}\bigg(1 + \frac{p}{\epsilon}\bigg)
	\bigg(1 + \frac{4\pi p r^3}{m}\bigg)\bigg(1 -
	\frac{2m}{r}\bigg)^{-1},
\\ \nonumber \\
&&\frac{dm}{dr} = 4 \pi r^2 \epsilon\\ \nonumber\\
&&\frac{d\nu}{dr} = - \frac{2}{\epsilon + p} \frac{dp}{dr},\;\;\;\;\; e^{\lambda(r)}\coloneqq\left[1-\frac{2m(r)}{r} \right]^{-1},
\label{tov3}
\end{eqnarray}
where $p$ is the pressure, $\epsilon$ is the energy density and $m$ the gravitational mass at the radial distance $r$. The $\lambda$ and $\nu$ functions are related to background metric, assumed to be given by
\begin{equation}
\label{dsz_tov}
ds^{2}=-e^{ \nu(r)} dt^{2} + e^{ \lambda(r)} dr^{2} + r^{2}(d\theta^{2}+\sin^{2}{\theta}d\phi^{2}).
\end{equation}
Tidal deformations of stars are taken to be related to nonradial perturbations with zero frequency. We work within the Regge-Wheeler gauge \citep{1957PhRv..108.1063R}, in which case (even parity) metric perturbations ($h_{ab}=\delta g_{ab}$) are diagonal and defined as
\begin{eqnarray}
h_{ab} &=& \mbox{diag}[H_0(r) e^{\nu(r)}, H_2(r)e^{\lambda(r)}, r^2 k(r),\nonumber \\ && r^2\sin^2\theta k(r)]Y^m_l(\theta,\phi)\label{hab},
\end{eqnarray}
where $(H_0,H_2,k)$ are functions to be determined by the Einstein field equations and $Y^m_l$ are the spherical harmonics. Without any loss of generality, and due to the spherically symmetric background, we work with $m=0$ and constrain our analysis to quadrupole deformations $(l=2)$.

The fluid perturbations are assumed to be given by \citep{2011PhRvD..84j3006P}
\begin{equation}
    \xi^r=\frac{W(r)}{r}P_2;\;\; \xi^{\theta}=\frac{V(r)}{r^2}\frac{d P_2}{d\theta};\;\;\xi^{\phi}=0\label{cont_pert},
\end{equation}
where $P_2(\theta)\propto Y^0_2$ is the Legendre polynomial of second order and $W(r)$ and $V(r)$ are functions to be determined by the field equations and conservation laws. Covariant components of the above perturbations are obtained readily through $\xi_a=g_{ab}\xi^b$. Note that there is no time dependence in the terms of Eq. \eqref{cont_pert}, meaning that only $u_t$ is different from zero. From the usual condition $u_au^a=-1$ and Eq. \eqref{hab}, we then get
\begin{equation}
    u_t= u_t^{0}+ \delta u_t= e^{\frac{\nu}{2}}\left(1 - \frac{1}{2} H_0 P_2\right)\label{utcov},
\end{equation}
where $\delta$ denotes the Eulerian perturbation.

We also take the baryon number conservation into account. It geometrically implies that the Lagrangian change of the baryon number density is given by \citep{2011PhRvD..84j3006P,2019CQGra..36j5004A}
\begin{equation}
\Delta n =  \delta n + \xi^a n_{;a}= -\frac{n}{2}{\cal P}^{ab}\Delta g_{ab}\label{Delta_nb},
\end{equation}
where the semicolon $(;)$ represents the covariant derivative \citep{1975ctf..book.....L} and
${\cal P}_{ab}\coloneqq g_{ab}+u_au_b$ is the projector onto the orthogonal directions of the four-velocity $u^a$. In addition,
\begin{eqnarray}
\Delta g_{cd}&=& h_{cd} + \xi_{c;d}+ \xi_{d;c}\nonumber\\
&=&h_{cd} + \partial_c\xi_d + \partial_d\xi_c-2\Gamma^a_{cd}\xi_a\label{Delta_metric},
\end{eqnarray}
with $\Gamma^a_{cd}$ the usual Christoffel symbols (connection coefficients) \citep{1975ctf..book.....L}. The Eulerian change of the energy momentum tensor of the fluid is assumed to be given by \citep{2011PhRvD..84j3006P,2019CQGra..36j5004A}
\begin{equation}
\delta T^b_a= (\delta T^b_a)_{\mbox{perf}} + \delta \Pi^b_a \label{delta_T_ab_total},
\end{equation}
where the first term is the perturbation of the energy momentum tensor of the perfect-fluid [$(\delta T^b_a)_{\mbox{perf}}=\mbox{diag} [-\delta \epsilon(r), \delta p(r),\delta p(r),\delta p(r)]P_2(\theta)$] and the second term represents the contribution from shear stresses \citep{2011PhRvD..84j3006P,2019CQGra..36j5004A}
\begin{equation}
\delta \Pi^{b}_a= -\tilde{\mu} \left({\cal P}^c_a{\cal P}^{db}-\frac{1}{3}{\cal P}_a^b{\cal P}^{cd}\right)\Delta g_{cd} \label{delta_pi},
\end{equation}
with $\tilde{\mu}$ the shear modulus. We note that this geometric formalism assumes perturbations with respect to an unstrained background configuration. For further details, see \citet{2019CQGra..36j5004A} and references therein.

As stressed by \citet{2011PhRvD..84j3006P}, one can check that for time-independent perturbations $\delta \Pi^b_t=0$. From Eqs. \eqref{delta_pi}, \eqref{Delta_metric} and \eqref{Delta_nb}, it also follows that
\begin{equation}
    \delta \Pi^r_r\coloneqq \delta\tilde{\Pi}^r_r P_2=\frac{2\tilde{\mu} P_2}{3r^2}[r^2(k-H_2)+(4-r\lambda')W-2rW'-6V]\label{deltaP11},
\end{equation}
where primes indicate radial derivatives. Also,
\begin{eqnarray}
    \delta \Pi^{\theta}_{\theta}&=& \frac{\tilde{\mu} }{3r^2}[\{-r^2(k-H_2)-(4-r\lambda')W + 2rW'\} P_2\nonumber \\ &+& 6V(2P_2-1)]\label{deltaP22},
\end{eqnarray}
\begin{equation}
    \delta \Pi^{\phi}_{\phi}= \frac{\tilde{\mu} }{3r^2}[\{-r^2(k-H_2)-(4-r\lambda')W + 2rW'\}P_2+ 6V]\label{deltaP33},
\end{equation}
\begin{equation}
    \delta \Pi^{\theta}_r\coloneqq \delta \tilde{\Pi}^{\theta}_r\frac{dP_2}{d\theta}= -\frac{\tilde{\mu} }{r^3}\frac{dP_2}{d\theta}[e^{\lambda} W -2V +rV']\label{deltaP21}.
\end{equation}
As expected, from Eqs. \eqref{deltaP11}--\eqref{deltaP33}, the trace of $\delta \Pi^a_b$ vanishes. Following \citet{2011PhRvD..84j3006P}, for future reference, we define
\begin{equation}
    T_{\theta}(r)\coloneqq -2\tilde{\mu} [e^{\lambda} W -2V +rV']\label{T2}
\end{equation}
and
\begin{equation}
    T_r(r)\coloneqq \frac{2}{3}\tilde{\mu} [r^2(k-H_2)+(4-r\lambda')W-2rW'-6V]\label{T1},
\end{equation}
related to the polar and radial components of the traction in the star ($T_{ab}n^b$, where $n^b$ is the normal four-vector to a given hypersurface), respectively, which helps us to apply the relevant boundary conditions.

Now we write down the Einstein equations relevant to our analysis. When one subtracts the $[\theta\theta]$ component of the Einstein equations from the $[\phi\phi]$ component, with the help of Eqs. \eqref{deltaP22} and \eqref{deltaP33}, it follows that
\begin{equation}
    H_2(r)= H_0(r) + 32\pi \tilde{\mu}V(r)\label{H2},
\end{equation}
confirming that solid stars spoil the equality between $H_2$ and $H_0$ \citep{2011PhRvD..84j3006P,2015PhRvD..92f3009K}, which holds for perfect fluid stars \citep{2008ApJ...677.1216H,2010PhRvD..81l3016H}. The $[r\theta]$ component of the Einstein equations tells us that
\begin{equation}
    k'(r)= \frac{8\pi}{r}(4\tilde{\mu} V-T_{\theta})+H_0'+H_0\nu'+16\pi\tilde{\mu}V\nu'\label{kprime}.
\end{equation}
The sum of $[\theta\theta]$ and $[\phi\phi]$ components, with the use of Eqs.~\eqref{deltaP22}, \eqref{deltaP33}, \eqref{H2} and \eqref{kprime} and the background equations, leads to
\begin{eqnarray}
    \delta p &=& \frac{T_r}{2r^2}+ \frac{e^{-\lambda}}{4 r^2}\{T_{\theta}[r(\lambda'-\nu')-2 ] -16e^{\lambda}\tilde{\mu}V -2 r T_{\theta}'\} + \nonumber \\
    &+& \frac{e^{-\lambda}}{16\pi r}H_0(\lambda'+\nu')\label{deltap},
\end{eqnarray}
which determines the radial part of the Eulerian change of the pressure. The $[rr]$ component of the Einstein equations can be simplified with Eqs. \eqref{H2} and \eqref{kprime}, and implies another relation for $k(r)$,
\begin{eqnarray}
   4e^{\lambda} k&=&H_0[6e^{\lambda} -2 -r(\lambda'+\nu') +(r\nu')^2] + + r^2H_0'\nu' + \nonumber\\
   &+& 8\pi e^{\lambda}[8 \tilde{\mu}V - 3T_{r}+ 2e^{-\lambda}\tilde{\mu}r^2 V (\nu')^2]- \nonumber\\
   &-&4\pi T_{\theta}[2+r(\lambda'+\nu')] + 8\pi r T_{\theta}' \label{k-equation-2}
\end{eqnarray}
Another relevant equation is the trace of the Einstein equations. By subtracting the $[tt]$ component from $[rr]+[\theta\theta]+ [\phi\phi]$, one gets
\begin{widetext}
\begin{eqnarray}
   16\pi r^2 e^{\lambda}(\delta \epsilon + 3\delta p) = &-&2H_0[2-8e^{\lambda}+r(\lambda' + 3\nu')-r^2(\nu')^2]-
   H_0'r[4-r(\lambda'-\nu')] -2r^2 H_0''- 16\pi r T_{\theta}\nu' +\nonumber\\
   &+&32\pi\tilde{\mu} V[r^2(\nu')^2-2r(\lambda' + 2\nu')+ 4e^{\lambda}-4]
   -32\pi r^2\nu'(\tilde{\mu}V)',
   \label{trace-einstein-equations}
\end{eqnarray}
\end{widetext}
where we recall that $\delta \epsilon=\delta \epsilon(r)$ is the radial part of the Eulerian change of the energy density of the star. For completeness, the $[tt]$ component of the Einstein equations implies that
\begin{eqnarray}
     8\pi r^2\delta\epsilon &=& -e^{-\lambda}r^2k'' - e^{-\lambda}\left(3-\frac{1}{2}r\lambda' \right) rk' + 2k \nonumber\\ &+& e^{-\lambda}r H_2' + \left[3 + e^{-\lambda}(1-r\lambda') \right]H_2\label{kpprime}.
\end{eqnarray}

Consistency with the case where perturbations have zero frequency is only obtained when the NS EOS is barotropic \citep{2011PhRvD..84j3006P}, that is, when $p=p(\epsilon)$, which results in
 \begin{equation}
     \delta\epsilon= \left(\frac{\partial \epsilon}{\partial p}\right)\delta p \equiv \frac{1}{c_s^2}\delta p\label{deltap-deltarho},
 \end{equation}
where $c_s^2$ is the adiabatic speed of the sound.

From  thermodynamical considerations, one can easily show that \citep{2011PhRvD..84j3006P}
\begin{equation}
    \Delta\epsilon \equiv \left(\delta \epsilon+ \frac{W}{rc_s^2}p'\right)P_2= (p+\epsilon) \frac{\Delta n}{n}\label{lagrangian-change-density}.
\end{equation}
From Eqs. \eqref{hab}, \eqref{cont_pert},  \eqref{Delta_nb} and \eqref{Delta_metric}, it follows that
\begin{eqnarray}
    \Delta n&=& -\frac{n}{2r^2} \left[ r^2(H_2+2k)+ W(r\lambda' + 2) \right. \nonumber \\ &+& \left. 2W'r -12 V \right]P_2\label{Delta-n}.
\end{eqnarray}
When the above equation is combined with Eqs. \eqref{deltap-deltarho} and \eqref{lagrangian-change-density}, one gets an additional equation for $\delta p$, namely,
\begin{eqnarray}
    \delta p&=& -c_s^2\frac{p+\epsilon}{2r^2} \left[ r^2(H_2+2k)+ W\left( r\lambda' + 2 -\frac{r\nu'}{c_s^2}\right)\right. \nonumber\\ &+&\left. 2W'r -12 V \right].\label{delta-p-baritropic}
\end{eqnarray}

Other equations, e.g. associated with the components of $(T^a_b)_{;a}=0$, may also be deduced. They are the,  usually not obvious, consequence of the field equations, hence it is enlightening to write them down. From $(T^a_r)_{;a}=0$, it follows that
\begin{eqnarray}
    \delta {\Pi}_r^r\left(\frac{6}{r}+\nu' \right) - (\epsilon + p)H_0'P_2 + 2(\delta p P_2 + \delta {\Pi}^r_r)'+ \nonumber\\
    (\delta p + \delta \epsilon)P_2\nu'-12\delta {\tilde\Pi}^{\theta}_rP_2 + 2p' + (p+\epsilon)\nu'=0\label{div_r}.
\end{eqnarray}
Consider now $(T^a_{\theta})_{;a}=0$, which implies that
\begin{eqnarray}
    &2&\tilde{\mu}[r^2(H_2-k)+18V +(r\lambda'-4)W + 2r W']+ \nonumber \\ &3&r^2\left\{2\delta p -H_0(p+\epsilon)+e^{-\lambda}r^2\left[2 (\delta \tilde{\Pi}^{\theta}_r)'+\delta \tilde{\Pi}^{\theta}_r \left(\frac{8}{r}-\lambda' + \nu'\right)\right] \right\}\nonumber\\ &=&0\label{div_theta}.
\end{eqnarray}
This equation reduces to $\delta p = H_0(\epsilon + p)/2$ when $\tilde{\mu}=0$ (perfect-fluid case). It is in full agreement with Eq.~\eqref{deltap} because $8\pi r e^{\lambda} (p+\epsilon)= (\lambda'+\nu')$. These equations will be useful when we deduce the appropriate boundary conditions to the perturbations (Sec. \ref{boundary-conditions}).

\section{Dense matter models}
\label{model}
In this first approach to hybrid stars, we assume a simplified model that contains some aspects of more realistic systems. For the quark phase, we mostly assume the Bag-like EOS \cite{2005ApJ...629..969A}, such that the grand thermodynamic potential is given by
\begin{equation}
\Omega =-p=  -\frac{3}{4\pi^2} a_4 \mu^4 + \frac{3}{4\pi^2}a_2\mu^2 + B
\label{grand_potential},
\end{equation}
where $\mu \equiv (\mu_u+\mu_s+\mu_d)/3$ is the averaged quark chemical potential, $a_4$ is a constant accounting for the strong interactions between quarks, $a_2$ is another constant which takes into account quark finite masses, quark pairing (color superconductivity), etc., and $B$ is a third constant encompassing the nontrivial vacuum of QCD \citep{2005ApJ...629..969A}. In the above expression, we have ignored the grand thermodynamic potential of electrons, $\Omega_e$, because it is usually much smaller than that of the quarks \citep{2018ApJ...860...12P}. From Eq. \eqref{grand_potential}, one can easily obtain other relevant thermodynamic quantities, such as the baryon number density $n_b\doteq -1/3(\partial \Omega/\partial \mu)$ and $\epsilon= 3\mu n_b-p$. From the above relations, one could isolate $\mu$ and get the following EOS \citep{2018ApJ...860...12P}
\begin{equation}
p(\epsilon)= \tfrac{1}{3}(\epsilon - 4B)-\frac{a_2^2}{12\pi^2a_4}\left[1+\sqrt{1+\frac{16\pi^2a_4}{a_2^2}(\epsilon -B)} \right],\label{eos_qm}
\end{equation}
which generalizes the usual bag model.
For the hadronic phase we consider either an effective polytropic EOS of the form $p=K\epsilon^2$ ($n=1$ of $p=K\epsilon^{1+\frac{1}{n}}$, $K=100$ km$^2$ \citep{2011PhRvD..84j3006P}), or work with hybrid star models involving the SLy4 EOS \citep{2001A&A...380..151D,2019A&A...622A.174S} and a relativistic mean field theory using the NL3 parameterization \citep{2018ApJ...860...12P}, connected to the BPS EOS \citep{1971ApJ...170..299B} for lower densities. For the model containing the SLy4 EOS, following \cite{2019A&A...622A.174S}, we assume that it is connected to a relativistic polytropic EOS \citep{Tooper1965} at a baryon density $n_0$ (freely chosen) with pressure $p=\kappa_{ef} n_b^{\gamma}$ and mass-energy density $\epsilon(n_b) = p(n_b)/(\gamma-1) + n_bmc^2$, where $\kappa_{ef}$, $\gamma$ and $m_b$ are free parameters ($m_b$ denoting the mass of the baryon in that phase, and $n_b$ the baryon density), and extends up to $n_1$ (transition baryon density of choice), where there is a sharp phase transition with a given energy density jump (free parameter). For baryon densities larger than $n_1$ we assume a quark phase described by a simple approximation to the bag model \citep{Zdunik2000}, in the form of $p=\alpha(\epsilon-\epsilon_*)$, with $\alpha$ a constant representing the square of the speed of sound, and $\epsilon_*$ a free parameter, denoting the density of the quark phase at zero pressure. Continuity of the pressure and specification of the energy density at the hadronic-quark interface uniquely fixes $\epsilon_{*}$. For the particular models we will use, we set $\gamma=4.5$, $n_1=0.335$ fm$^{-3}$ (above twice the nuclear saturation density, $\epsilon_{\mbox{sat}}=2.7\times 10^{14}$ g cm$^{-3}$) and $\alpha=1$ to also investigate properties of stiff quark matter (see Figure 9 of \citealt{2019A&A...622A.174S} and the text therein for more details).

Table \ref{ta_models} summarizes the main aspects of the hybrid star models we use in this work. We assume the Maxwell construction (constancy of the pressure and the chemical potential at a sharp phase-splitting surface) \citep{2018arXiv180901116B} to determine the aspects of background hybrid stars. For the HS1 and HS3 models, we take the density jump at the quark-hadronic phase interface as a free parameter. In the case of the HS1 model in particular, given a $\eta \coloneqq \epsilon_q/\epsilon_h-1$ (possible values for $\eta$ are in the range $0-2.0$), where $\epsilon_q$ is the density at the top of the quark phase and $\epsilon_h$ the density at the bottom of the hadronic phase, from $p_q=p_h$ at the quark-hadronic phase transition surface one can easily find either $\epsilon_q$ or $\epsilon_h$, which then allows for the determination of the phase transition pressure and the chemical potential. For the HS2 model, $\eta$ is uniquely found by the Maxwell construction.

\begin{center}
\begin{table}
[htbp] 
\caption{Main aspects of the hybrid star models used in our work.}
\begin{ruledtabular}
{\begin{tabular}{@{}c|cccc@{}}
Hybrid& Quark Model & $\eta$ & Hadronic \\
Model&$\left(\frac{B^{\frac{1}{4}}}{\mbox{MeV}},a_4, \frac{a_2^{\frac{1}{2}}}{\mbox{MeV}}\right)$ & $\left(\frac{\epsilon_q}{\epsilon_h}-1\right)$& Model\\ \hline
HS1&$(137,0.40,100)$ & free & Polyt. (n=1)\\
HS2&$(140,0.55,100)$ & $\approx$ 0.45 & BPS+NL3\\
HS3& Bag with $c_s^2=1$ &free & SLy4+Polyt. ($\gamma=4.5$)\\
\end{tabular} \label{ta_models}}
\end{ruledtabular}
\\
\end{table}
\end{center}

We take the quark phase to be a perfect fluid, even though it is also theoretically possible that the quark phase is elastic \citep{2007PhRvD..76g4026M}. Unless otherwise stated, we assume that the hadronic phase is elastic only for densities below $\epsilon~\approx~2\times 10^{14}$\ g cm$^{-3}$ ($\approx 2/3$ of the nuclear saturation density) \citep{2011PhRvC..83d5810D}. For the shear modulus, we take the simple linear model \citep{2011PhRvD..84j3006P,2008LRR....11...10C}
\begin{equation}
    \tilde{\mu}(r)= \kappa p(r) + \tilde{\mu}_0\label{shear_modulus_model},
\end{equation}
where $\kappa$ and $\tilde{\mu}_0$ are given constants.

Naturally, our models (for the EOSs and the shear modulus) are simplistic but they are expected to give us the main aspects hybrid stars with elastic hadronic regions (solid crusts) should present, which could be used as input for more accurate descriptions. We assume that the solid crust ends at the density $10^7$ g cm$^{-3}$, where the envelope (or the ocean) of a NS is supposed to start \citep{2011PhRvD..84j3006P,2007ASSL..326.....H}.\footnote{The density where the crust melts depends on the temperature (see, e.g., \citealt{2020arXiv200305449G}) and is believed to be in the range of $10^6-10^8$~g cm$^{-3}$ for cold NSs \citep{2007ASSL..326.....H}. The precise value of this density is not crucial for tidal deformation calculations because the low density region of a star does not have a significant impact on its quadrupole moment.}

\section{Boundary conditions}
\label{boundary-conditions}

In order to integrate the Einstein equations and obtain the tidal deformations, one needs to impose appropriate boundary conditions at the center, the liquid-elastic interfaces
and the surface of the star. We describe them here.

At the center of a hybrid star, we impose that all functions are regular. In other words, we expand a given function $F$ as \citep{2015PhRvD..92f3009K}
\begin{equation}
    F(r)=F_0 + \frac{1}{2}F_2r^2+ {\cal O}(r^4)\label{origin-expansion},
\end{equation}
where the coefficients $F_0$ and $F_2$ are to be determined from the field equations (the symmetry of the field equations render first order terms null,  \citealt{2015PhRvD..92f3009K}). One also needs to expand the background quantities $\epsilon$, $p$, $\lambda$ and $\nu$ in the same fashion. One can always express second order corrections to the background ($F_2$ in Eq. \ref{origin-expansion}) in terms of the solutions to the TOV equations; for the explicit expressions, see \citet{2015PhRvD..92f3009K}. Naturally, when working numerically, one could directly find second order corrections to the background in terms of their second derivatives. When zeroth and second order coefficients to the perturbations are found, one has the initial conditions to start the core integration.

For a liquid-elastic interface, boundary conditions could be easily found from aspects of the extrinsic curvature. We restrict ourselves to the case of no surface degrees of freedom (continuity of the extrinsic curvature, see \citealt{2004reto.book.....P}).
We assume that a liquid-elastic interface is described by $\Psi=r-R_{le}-\xi^r=0$, where $R_{le}$ is the background liquid-elastic phase transition radius.\footnote{If the density at the base of the hadronic phase is larger than the critical density marking the onset of elasticity, then a hybrid star should have three main (internal) interfaces: (i) one separating the top of the liquid quark phase from the bottom of the liquid hadronic phase, (ii) another one between the top of the liquid hadronic phase and the bottom of the solid crust and (iii) a third one splitting the top of the solid crust from the bottom of the liquid ocean. These phase-splitting surfaces are described in the same physical way with regard to boundary conditions (continuity of the induced geometry and extrinsic curvature; see \citealt{2020arXiv200305449G}). Therefore, in addition to a perfect fluid-perfect fluid (or liquid-liquid) phase transition, we just need to determine the boundary conditions for perturbations at a liquid-elastic interface.} Standard calculations for the extrinsic curvature ($K^{a}_b$, $\{a,b\}=t,\theta,\varphi$) \citep{1990MNRAS.245...82F,2020arXiv200305449G} lead to the following nontrivial components at $r=R_{le}$:
\begin{equation}
    \delta K^t_t=-4e^{-\frac{1}{2}\lambda (R_{le})}P_2\left( H_0'+\frac{1}{2}H_2\nu'-\frac{W}{r}\nu''+\frac{1}{2}\frac{W}{r}\nu'\lambda'\right)_{r=R_{le}}\label{k00}
\end{equation}
and
\begin{equation}
    \delta K^{\theta}_{\theta}=\delta K^{\phi}_{\phi}=-\frac{4}{R_{le}^2}e^{-\frac{1}{2}\lambda(R_{le})}P_2\left( rH_2-r^2k'+W\lambda'\right)_{r=R_{le}}.\label{k11}
\end{equation}
The continuity of the extrinsic curvature implies the terms in parenthesis in the above equations have null jumps. (Note that the background components of the extrinsic curvature \citep{2020arXiv200305449G} are automatically continuous when the Maxwell construction is assumed.) From Eq. \eqref{k00}, one readily finds that
\begin{equation}
    [H_0']^+_-= 8\pi e^{\lambda(R_{le})}\frac{W}{R_{le}}[\epsilon]^+_- -16\pi\mu^+ V^+\nu'\label{jumpofH00},
\end{equation}
where we have made use of $[\nu''-\nu'\lambda'/2]^+_-= 8\pi e^{\lambda}[\epsilon]^+_-$ (see e.g. \citealt{1975ctf..book.....L} for the precise expression for $\nu''-\nu'\lambda'/2$) and Eq.~\eqref{H2}. In addition, ``+''~(``-'') relates to the bottom of an elastic phase (top of the liquid phase), which implies that $\tilde\mu^-=0$. The continuity of the induced metric at $r=R_{le}$ also leads to $[W]^+_-=0$ (see \citealt{2020arXiv200305449G} for further details). From Eq. \eqref{cont_pert}, it follows that $[\xi^r]^+_-=0$. It turns out the null discontinuity of $K^{\theta}_{\theta}$ does not add a new jump condition. This could be readily seen from Eqs. \eqref{kprime} and \eqref{jumpofH00} and the fact that $[\lambda']^+_-= 8\pi r e^{\lambda}[\epsilon]^+_-$.

The interface condition of other functions could be more easily found when perturbation equations are promoted to distributions.\footnote{For further details on distributions, see \cite{2004reto.book.....P}. In summary, one assumes that a physical quantity $X(r)$ is decomposed as $X=X_-\theta(R_t-r)+X_+\theta(r-R_t)$, where $R_t$ is the radius where a sharp phase transition takes place and $\theta$ is the Heaviside (step) function. Boundary conditions are obtained by collecting the Dirac delta terms in the equations.} For an alternative and equivalent results based on the continuity of the induced metric and extrinsic curvature, see \citep{2020arXiv200305449G}. Given that we have second derivatives of $H_0$ [see Eq. \eqref{trace-einstein-equations}], the promotion of the Einstein equations to distributions would only make sense if the jump of $H_0$ at $R_{le}$ is null, i.e., $[H_0]^+_-=0$. Eq. \eqref{div_r} then shows us that $[\delta p + p'W/r + \delta\tilde{\Pi}^r_r]^+_-=0$, which, from Eqs. \eqref{T1} and \eqref{deltaP11}, implies that $[T_r+r^2\delta p+r W p']^+_-=0$. From Eq. \eqref{div_theta}, we have that $[\delta \tilde{\Pi}^r_{\theta}]^+_-=0$, which, due to Eq. \eqref{T2}, leads to $[T_{\theta}]^+_-=0$. From the above jump conditions, one has $[k]^+_-=0$, as can be easily seen from Eq. \eqref{kprime} when promoted to a distribution.

At the surface of the star ($r=R$) and outside it, the reasoning of \cite{2008ApJ...677.1216H,2009PhRvD..80h4035D,2010PhRvD..81l3016H} ensue. In other words, if $H_0'$ is continuous at the star's surface, one can obtain the Love number $k_2$ in terms of $y\equiv RH_0'(R)/H_0(R)$ as \citep{2008ApJ...677.1216H,2009PhRvD..80h4035D,2010PhRvD..81l3016H,2011PhRvD..84j3006P}
\begin{widetext}
\begin{equation}
    k_2=\frac{8C^5(1-2C)^2[2+2C(y-1)-y]}{5\{2C[6-3y+3C(5y-8)]+4C^3[13-11y +C(3y-2)+2C^2(1+y)]+3(1-2C)^2[2-y+2C(y-1)]\ln(1-2C)\}}\label{love-number},
\end{equation}
\end{widetext}
where $C\equiv M/R$ is the compactness of the background hybrid star. If $H_0'$ is not continuous at the stellar surface, for instance due to a density discontinuity or a non-zero shear modulus, as in Eq. \eqref{jumpofH00}, then $k_2$ also changes due to the discontinuity of $y$ \citep{2009PhRvD..80h4035D,2010PhRvD..81l3016H}. From $k_2$, one can construct the physically relevant tidal deformation $\Lambda\doteq 2/3(M/R)^{-5}k_2$. We will focus mainly on this quantity for the ease of comparison with the GW constraints.

We turn now to a simplification of boundary conditions at a liquid-elastic interface. We recall that in this case, ``+'' (``-'' ) relates to the elastic (liquid) phase.
It is easy to show that the condition $[T_r+r^2\delta p + r W p']^+_-=0$ implies
\begin{widetext}
\begin{equation}
    V^+= \frac{1}{16\pi\tilde\mu^+R_{le}^2\nu'^2}\left[ 4 e^{\lambda} \{ k_{le}+4\pi R_{le} W_{le} (p'_--p'_+)\} -R_{le}^2H_0'^+\nu' + H_0(2+2e^{\lambda}\{ 4\pi R_{le}^2(p^-+\epsilon^-)\}-R_{le}^2\nu'^2)\right]\label{Vplus},
\end{equation}
\end{widetext}
with $k_{le}\coloneqq k^-$ (Eq. \ref{k-equation-2} evaluated at the top of the perfect-fluid phase where $\tilde\mu=0$) and $ W^+=W^-\coloneqq W_{le}$. In this equation it is tacit that $\nu'\equiv\nu'(R_{le})$ and $\lambda\equiv\lambda(R_{le})$. The condition $[T_{\theta}]^+_-=0$
and Eqs. \eqref{Vplus} and \eqref{jumpofH00} actually lead to an algebraic system of equations to be solved for $H_0'^+$ and $V^+$, for instance, and its not difficult to show that its solution is
\begin{eqnarray}
  W_{le}&=&\mbox{unconstrained},\nonumber\\
    V^+&=&\frac{1}{2}\left(R_{le}V'_+ + e^{\lambda(R_{le})}W_{le} \right),\label{cc-bou}\\
     H_0'^+&=&H_0'^- - 16\pi V^+R_{le}\tilde\mu^+\nu'(R_{le})+\frac{8\pi}{R_{le}}e^{\lambda(R_{le})} [\epsilon]^+_-W_{le}\nonumber.
\end{eqnarray}
For the solid crust-liquid envelope interface ($r=R_{ce}$), similarly to Eq. \eqref{cc-bou}, the relevant boundary conditions are:
\begin{eqnarray}
 T_r^-&=& R_{ce}^2[\delta p]^+_ - + W_{ce}R_{ce}[p']^+_-,\nonumber\\
 T_{\theta}^-&=&0 \rightarrow V^-=\frac{1}{2}\left(R_{ce}V'_- + e^{\lambda(R_{ce})}W_{ce} \right),\label{crust_env-bou}\\
     H_0'^+&=&H_0'^- + 16\pi V^-R_{ce}\tilde\mu^-\nu'(R_{ce})+\frac{8\pi}{R_{ce}}e^{\lambda(R_{ce})} [\epsilon]^+_-W_{ce},\nonumber\\
     k^+&=&k^-\nonumber,
\end{eqnarray}
where now the ``-'' (``+'') now stands for the top of the solid crust (base of the ocean) and  $W^+=W^-\coloneqq W_{ce}=\mbox{unconstrained}$. Given that $k^+$, $\delta p^+$ are related to a perfect fluid, it is not difficult to show that the first, third and fourth equations from Eq. \eqref{crust_env-bou} result in only one nontrivial condition to be fulfilled.

In summary, we have two free parameters (e.g. $V_+'$ and $W_{le}$) at the base of the solid crust fulfilling Eqs. \eqref{cc-bou} and two constraints at the top of the solid crust, e.g., $V^-$ and $H_0'^-$, compatible with Eqs. \eqref{crust_env-bou}. Thus, the problem is well posed and its solution is unique.  For the ocean, the boundary conditions for the perfect-fluid tidal deformation integrations are the continuity of $H_0$ at the crust-envelope interface and $H_0'^+$ from Eq.~\eqref{crust_env-bou}. We assume in our calculations that the density jump at the interface between the top of the crust and the base of the envelope is negligible.\footnote{The assumption of the approximate crust-envelope density continuity along with the expected smallness of $\tilde{\mu}$ at be top of the crust and the thin ocean ($\lesssim 0.1\%$ of the star radius) imply that $y(R)$ will in general deviate little from $y(R_{ce})$. Given that $\Lambda$ depends non-linearly on $y$, the differences between $\Lambda(R)$ and $\Lambda(R_{ce})$ will in general be even smaller. Thus, in this case, one could stop tidal deformation integrations at the top of the crust.} The density jump at the interface separating the perfect-fluid hadronic phase from the solid crust is also taken as insignificant \citep{2000MNRAS.319..902U,2008LRR....11...10C}.

The constraints we have derived cover all sets of equations one might want to integrate in the solid crust. In our calculations, we choose to work with $(k,H_0,V,W)$. Due to the re-scaling invariance of the equations, solutions for the perturbations should depend on an arbitrary amplitude which naturally does not affect physical quantities like the Love numbers.
Given that we have two arbitrary quantities at the base of the solid crust (e.g. $W$ and $V'$) to be fixed by two boundary conditions at the top of the crust, a shooting method should be used to integrate the equations in the elastic region. After the integration fulfilling all boundary conditions, one is just left to evaluate Eq.~\eqref{love-number}.

For completeness, we note that the constraints in Eqs.~ \eqref{cc-bou} (or Eqs. \ref{crust_env-bou}) are only meaningful if $\tilde\mu^+\neq 0$ ($\tilde\mu^-\neq 0$). If this is not the case, then boundary conditions valid for perfect-fluids should be used. More specifically, the continuity of the radial traction and the extrinsic curvature at a perfect fluid-perfect fluid (liquid-liquid) interface (at $r=R_{ll}$) imply that
\begin{eqnarray}
   &&W_{{ll}}=-\frac{[r(\delta p)_{{ll}}]^+_-}{[p_0']^+_-}= \left[\frac{H_0 r^3}{2(m+4\pi r^3 p)}e^{-\lambda} \right]_{r=R_{ll}},\nonumber\\
     &&(H_0'^+)_{{ll}}=(H_0'^-)_{ll} +{8\pi}e^{\lambda(R_{ll})}\frac{W_{{ll}}}{R_{ll}}[\epsilon]^+_-
     \label{perf-fluid-bound}.
\end{eqnarray}
We use this constraint on $(H_0')_{ll}$ for any tidal deformation calculation involving liquid-liquid interfaces in hybrid stars. In these cases, naturally, we solve the the perfect-fluid equation for $H_0$ as in \cite{2008ApJ...677.1216H}.

\section{Results}
\label{results}

In our analysis, we have taken $\kappa=1.5\times10^{-2}$ and $\tilde{\mu}_0=0$ (physically, $\tilde{\mu}_0\lesssim 10^{-25}$cm$^{-2}$).
These parameters are reasonable, given that shear moduli should be around $1\%$ of the crust pressure \citep{2008LRR....11...10C,2003ApJ...588..975C}. As a first cross-check of our equations, we have applied them to the case investigated in \cite{2011PhRvD..84j3006P}. We have worked with the coupled system of equations for $(H_0,k,W,V)$ and applied the standard 4$th$ order Runge-Kutta method. The number of steps used in the solid crust integration is $10^6$, and we have kept an accuracy of $12$ significant digits (and a much higher precision). The values obtained for  $k_2$ in the case of elastic crusts is smaller than the perfect fluid case, and relative changes were $10^{-6}-10^{-3}$, one order of magnitude smaller than the results quoted in \cite{2011PhRvD..84j3006P}. This is in full agreement with the approach of \cite{2020arXiv200305449G}. We have varied the value of $\tilde\mu$ in the solid crust and have found that indeed the solution to $H_0$ converges to the perfect fluid case when shear stresses go to zero. This shows that the system of equations we have chosen does not present numerical instabilities.

\begin{figure}[htbp]
\centering
\includegraphics[width=1.05\columnwidth,clip]{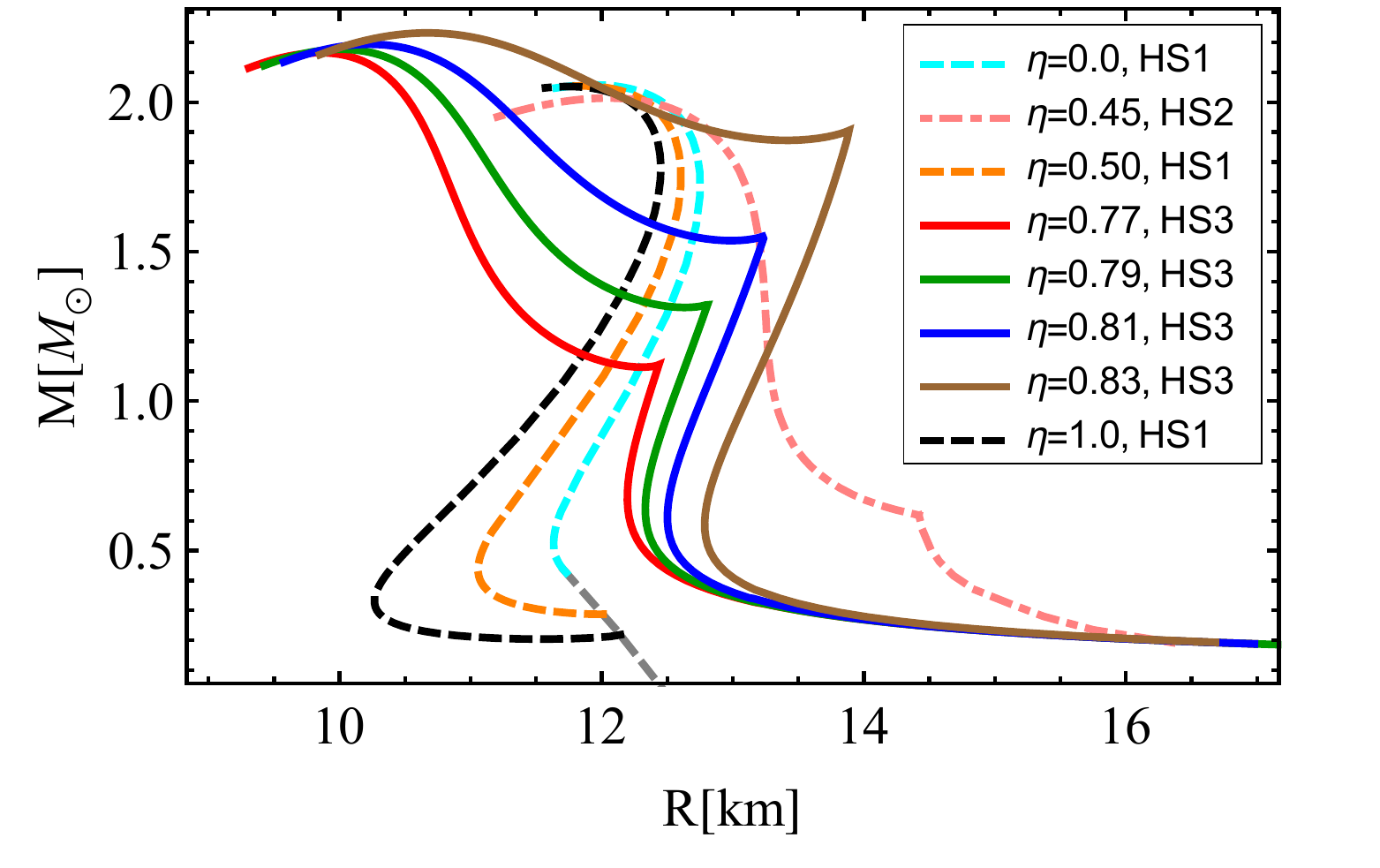}
\caption{Mass-radius $M(R)$ relations for the hybrid star models of Table \ref{ta_models} used in this work. The HS1 model is used as an example which is convenient for numerical calculation checks and general aspects of the tidal deformations of hybrid stars with solid crusts. The HS2 and HS3 models are somewhat more realistic.
From bottom to top of the $M(R)$ thick curves (from red to brown), we have chosen $n_0=(0.235,0.21,0.185,0.16)$ fm$^{-3}$, respectively. All models lead to different values of $\eta \equiv \epsilon_q/\epsilon_h-1$ (see the text for details), as shown for each curve. The  same can be said regarding the critical masses (the cusps of each curve) above which a quark phase appears. For all models, tidal deformations of $1.4$ M$_{\odot}$ perfect-fluid stars (with different $\eta$) are smaller than $\approx 650$.
}
\label{model1}
\end{figure}

We have solved the same system of equations for the models presented in Table \ref{ta_models}, taking some representative values for the free parameters.
Figure \ref{model1} shows some of the corresponding $M-R$ relations. We stress that the HS1 model is simply an example of a hybrid EOS and has been used because it is convenient for numerical checks and is expected to give us the main results of the problem of tidal deformations of hybrid stars with elastic crusts. We have chosen the HS1 parameters so that the largest masses of hybrid stars are above two solar masses and the tidal deformations of 1.4~M$_\odot$ stars is smaller than the upper limits found by LIGO/VIRGO \citep{2017PhRvL.119p1101A,2019PhRvX...9a1001A}. The same has been done for the parameters of the somewhat more realistic HS2 and HS3 models (see Table \ref{ta_models}).\footnote{The HS1 and HS3 EOSs are stiffer than the HS2 EOS, which allows us to (roughly) estimate upper limits to the relative tidal deformation changes.} The masses marking the appearance of the quark phase (the cusps of the $M(R)$ relations) also vary, reaching larger values for the HS3 EOSs \citep{2019A&A...622A.174S}. In our study, we focus on the cases where $\partial M/\partial \epsilon_c> 0$, with $\epsilon_c$ the central density of a family of solutions for an EOS. We do not enter into the details of the stability of hybrid stars with elastic phases, but rather assume that the conditions valid for perfect fluids are also valid here.

\begin{figure}[htbp]
\centering
\includegraphics[width=\hsize,clip]{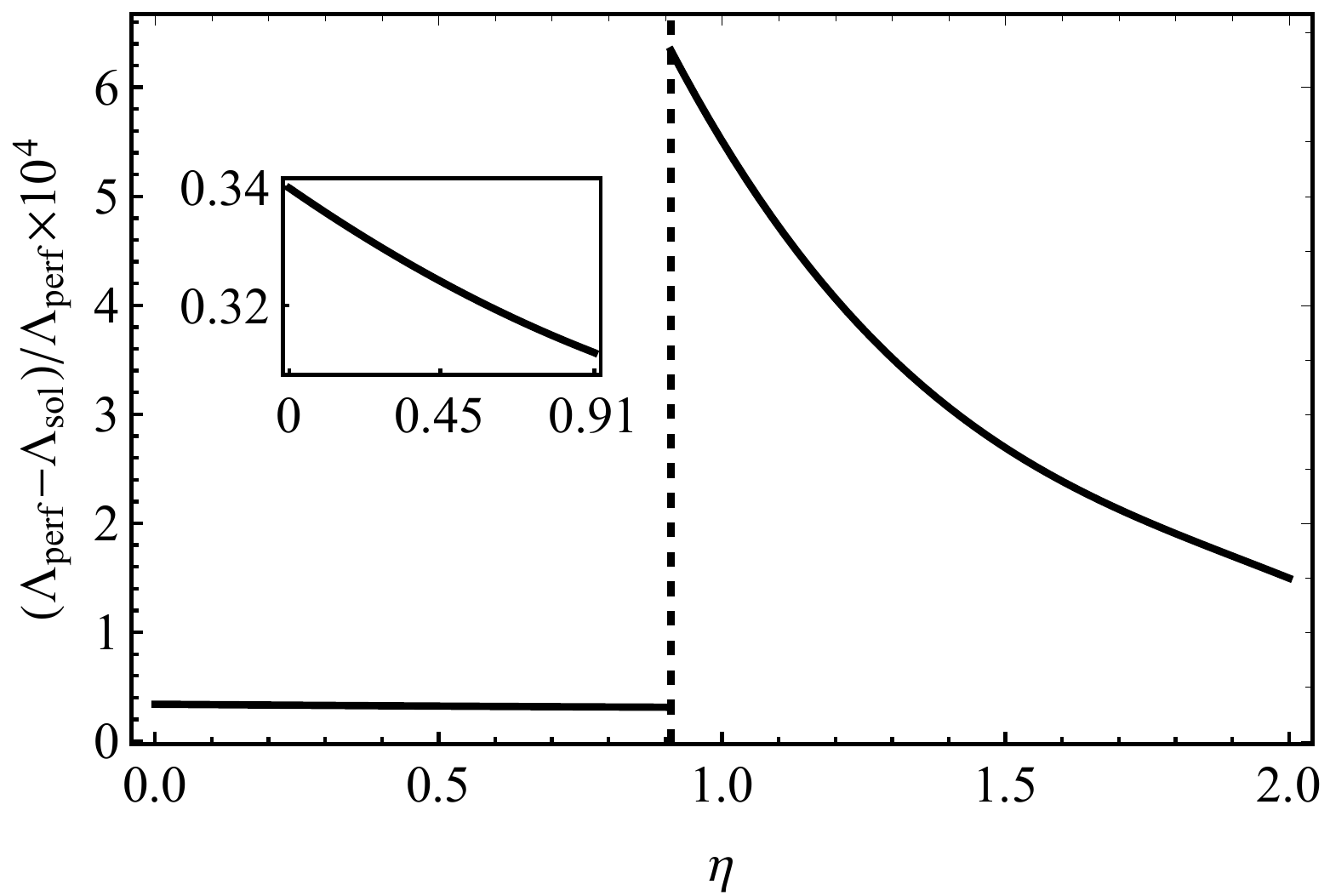}
\caption{Tidal deformations of 1.4 M$_{\odot}$ solid hybrid stars as a function of the density jump ($\eta$) for the HS1 EOS. For different $\eta$s, different central pressures have been chosen to have the same mass. For the HS1 model, $\eta\approx 0.91$ determines the critical density discontinuity above which the elastic crust directly touches the quark phase. For lower density jumps, there is a hadronic perfect-fluid layer separating the quark phase from the elastic hadronic phase. We have assumed that the density is approximately continuous at the interface separating the hadronic perfect-fluid from the hadronic elastic phase \citep{2000MNRAS.319..902U,2008LRR....11...10C}. This is the main reason for the discontinuous behavior of the relative change of the tidal deformation in this plot. For the model used, the quark phase covers a significant part of the star. In this case, tidal deformation changes due to an elastic phase should be small and around $10^{-3}-10^{-2}\%$.}
\label{tidal_def_1.4}
\end{figure}

Figure \ref{tidal_def_1.4} shows the expected relative tidal deformation changes [$(\Lambda_{\mbox{perf}}-\Lambda_{\mbox{sol}})/\Lambda_{\mbox{perf}}$] for 1.4M$_{\odot}$ hybrid stars ($\Lambda_{\mbox{perf}}^{1.4}\approx 640$) with large quark phases (${\sim}80\%$ of the star's radius) and small elastic crusts (${\sim}10\%$ of $R$). We use the HS1 model for these calculations, given its flexibility with density jumps. Relative changes are very small, of the order of $10^{-3}-10^{-2}\%$. We find a weak dependence of the relative tidal deformation change on the density discontinuity for values smaller than the critical density jump, above which the whole hadronic phase is elastic. For the parameters chosen, $\eta_{\mbox{crit}}\approx 0.91$. The sharp change at this density appears as the boundary conditions for tidal perturbations are changed due to the absence of a perfect fluid layer (of hadronic matter) separating the quark phase from the elastic crust. The above-mentioned weak dependence is mostly due to the fact that a density discontinuity is accompanied by a change of radius and mass of the star, which almost counteracts the tidal deformation changes.

Table \ref{ta1} shows some consequences (all satisfying the condition $\partial M/\partial \epsilon_c>0$) of the models of Table \ref{ta_models}, for instance their relative tidal deformation changes with respect to perfect-fluid hybrid stars. We have assumed that the shear modulus of all EOSs is given by Eq.~\eqref{shear_modulus_model} and we have taken $\kappa$ and $\tilde{\mu}_0$ the same as in the beginning of this section. We find that for all models the relative tidal deformation changes are of order of $2-4\%$ for solid hadronic phases larger than ${\sim}60\%$ of the star's radius. For the other cases, the changes are much smaller. We conclude that, in addition to the EOS itself, relative tidal deformation changes depend mostly on the sizes of the elastic phases. As expected, $(\Delta \Lambda/\Lambda_{\mbox{perf}})$ decreases with the decrease of the thickness of the elastic crust. We comment on aspects of the last column of Table \ref{ta1} in the following.

\begin{figure}[htbp]
\centering
\includegraphics[width=1.1\columnwidth,clip]{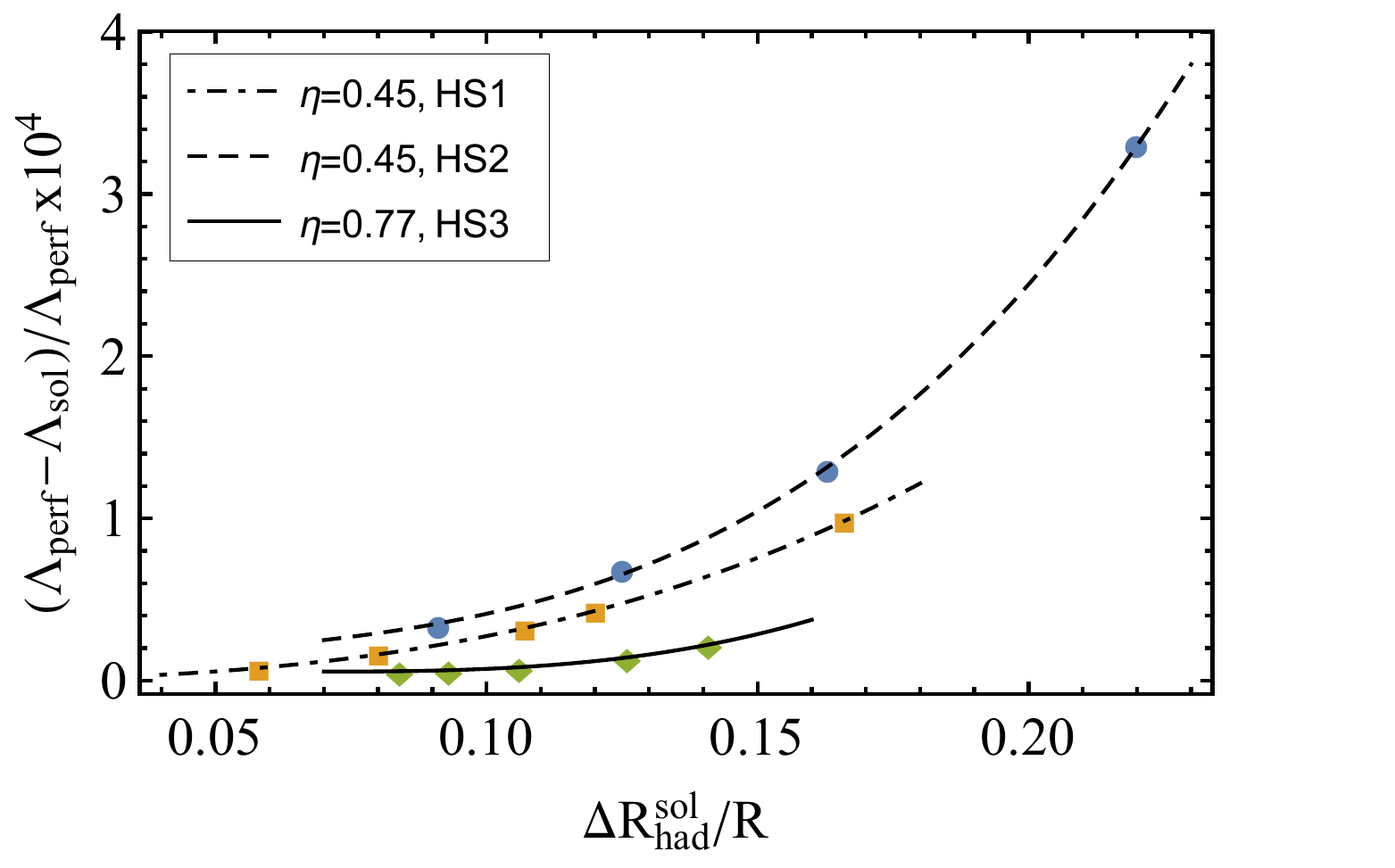}
\caption{Relative tidal deformation changes as a function of the relative thickness of the elastic hadronic phase for $\eta=0.45$ and $0.77$ of Table \ref{ta1}. Cubic fits ($ax+bx^2+cx^3$) describe well the relationship between relative tidal deformation changes and $\Delta R^{\mbox{sol}}_{\mbox{had}}/R$.}
\label{tidal_def_thickness}
\end{figure}

Figure \ref{tidal_def_thickness} shows some relative tidal deformation changes for selected $\eta$ of Table \ref{ta1} as a function of the thickness of the elastic hadronic phase. We find that (different) cubic fits (of the form $ax+bx^2+cx^3$) could describe $(\Delta \Lambda/\Lambda_{\mbox{perf}})$ as a function of $(\Delta R^{\mbox{sol}}_{\mbox{had}}/R)$ for all hybrid models. We note that the fits agree with the expected condition that $\Delta \Lambda\rightarrow 0$ when the thickness of the elastic crust goes to zero. We have also found similar dependencies of the relative tidal deformation change as a function of  $(1/C)\Delta R^{\mbox{sol}}_{\mbox{had}}/R$ for each model, where we recall that $C$ is the compactness.

Table \ref{ta1} also explores the cases where the elastic hadronic phase comprises a significant portion of a hybrid star. In order to consider this case, we have assumed a nonzero shear modulus (around $1-2\%$ of the local pressure) right from the bottom of the hadronic phase. The results for the relative tidal deformation changes are given in the last column of this table. At this level these results are merely indicative, but they are motivated by the presence of an elastic mixed phase in a hybrid star \citep{2013NuPhA.906...37S}. In this case, we find that very different models (for example the HS1, HS2 and HS3) roughly agree with the ``largest'' relative tidal deformation changes, which are also around $3-4\%$ when the solid hadronic region comprises over $75\%$ of the star's radius.

When realistic shear moduli are considered, the results for the tidal deformation changes might change. For instance, the SLy4 shear modulus \citep{2006MNRAS.373.1423H,2007MNRAS.375..261S,2008A&A...491..489Z} is around 1-5 times smaller than the one of Eq. \eqref{shear_modulus_model} in the range where the hadronic phase is elastic. This means, roughly speaking, that similar changes are expected for tidal deformation quantities we have calculated before. Indeed, numerical calculations with the SLy4 shear modulus for all the models of Table \ref{ta_models} lead to this level of decrease in $(\Delta \Lambda/\Lambda_{\mbox{perf}})$. Therefore, relative tidal deformation changes of hybrid NSs are not expected to be larger than $3-4\%$ of their perfect-fluid values.

\subsection{Phase-splitting surface quantities}
An interesting issue in hybrid stars is the possibility of inducing, due to perturbations, surface quantities on the phase-splitting interfaces (see, e.g., \citealt{2019ApJ...871...47P}). Surface quantities induced by perturbations could significantly change the boundary conditions, which, in turn, may change the tidal deformations of elastic stars. We show in this section, nonetheless, that the Einstein equations for quasi-static perturbations {\it do not allow} the induction of surface quantities at phase-splitting surfaces.

From the Darmois-Israel formalism (see, e.g., \citealt{2014PhRvD..90l3011P}), we know that surface quantities are in general necessary for the consistent match of two spacetimes, and are related to the presence of a surface energy-momentum tensor $S^a_b$ of the form \citep{2014PhRvD..90l3011P}
\begin{equation}
    S^a_b=-\frac{1}{8\pi}[K^a_b-\delta^a_b K]^+_-\label{Sab},
\end{equation}
where we recall that $K^a_b$ is the extrinsic curvature and $K=K^a_a=K^{ab}\bar{h}_{ab}$ ($\bar{h}_{ab}$ is the induced metric on the hypersurface splitting two given spacetimes) is its trace. From Eqs. \eqref{k00} and \eqref{k11}, it is easy to see that $S^a_b$ should be related to a perfect fluid, so that one could write
\begin{equation}
    S_{ab}=(\sigma + {\cal P})u_au_b + {\cal P}\bar{h}_{ab}\label{Sabperf},
\end{equation}
where
\begin{equation}
    \sigma = -\frac{1}{4\pi}[K^{\theta}_{\theta}]^+_-\label{sigma}
\end{equation}
and
\begin{equation}
    {\cal P}= \frac{1}{8\pi}[K^t_t+K^{\theta}_{\theta}]^+_-=\frac{1}{8\pi}[K^t_t]^+_- -\frac{\sigma}{2}.\label{surface_tension}
\end{equation}
The physical reason for the perfect fluid nature of the surface energy momentum tensor is the axial symmetry of the spacetimes (perturbed core and crust).

Equation~\eqref{kprime} allows us to easily find $[k']^+_-$, which, from Eq.~\eqref{sigma}, relates $[H_0']^+_-$ to $\sigma$. The discontinuity of $H_0'$, $[H_0']^+_-$, on the other hand, can be determined by Eqs. \eqref{surface_tension} and \eqref{k00}. When one solves the associated equation, it follows that ${\cal P}=0$. Since ${\cal P}$ is a monotonic function of $\sigma$ in ordinary cases, we have $\sigma=0$, too. Thus, the intrinsic curvature components should always be continuous at any phase-transition interface.

\section{Concluding remarks}
\label{discussion}

We have investigated tidal deformations of hybrid stars with elastic hadronic phases, building on previous works and extending them to the case of sharp phase transitions, which is a possibility for hybrid stars. In the case of a shear modulus around 1$\%$ of the hadronic pressure, stars with small elastic hadronic phases (large perfect-fluid quark phases) lead to negligible changes in $\Lambda$. However, there are cases where the tidal deformation changes with respect to perfect fluids could be as large as $2-4\%$. These cases relate to elastic hadronic regions larger than ${\sim}60\%$ of the stellar radius. This would imply, for instance, small quark phases and configurations with large enough density jumps at the interface separating the quark phase from the hadronic phase, or even small quark phases with lower phase transition pressures. One would expect these cases to maximize the relative tidal deformation changes because they are exactly the ones where a solid phase would comprise the largest fraction of the star.

Although ${\simeq}5\%$ changes in the tidal deformability will be difficult to register for current GW detectors, for signals emitted from typical distances of 100 Mpc, the issue of detecting elastic hadronic regions of stars may become relevant for future detectors, such as the Einstein Telescope \citep{2010CQGra..27s4002P}. Indeed, in the recent review of the Einstein's Telescope science  case \citep{2019arXiv191202622M}, the expected increase of sensitivity translates into more than one order of magnitude larger values for the signal-to-noise. In the limit of high signal-to-noise, Fisher-information matrix approximation suggests that the errors on the measured quantities scale inversely proportional to the signal-to-noise \citep{PhysRevD.98.124014,PhysRevD.95.104004,PhysRevD.77.042001}. For example, for a GW170817-like event observed by the Einstein Telescope, the errors of the measured component mass-weighted tidal deformability $\tilde{\Lambda}$ (equal to $300^{+420}_{-230}$ in the low-spin prior case, \citealt{2019PhRvX...9a1001A}) would be sufficiently small to allow for studies of the elastic properties.

We have also found that there is a weak dependence of the tidal deformations of solid stars on density jumps. The reason seems to be a competition between larger density discontinuities and smaller masses (radii) for stars with a given central density. However, a sudden change might appear whenever there is not a layer of perfect fluid separating the quark phase from the solid crust. This might happen when density jumps are large or the critical density favoring dense matter to arrange as a lattice is high enough. We recall that this critical density is not known (it is EOS dependent), but it is believed to be smaller than the nuclear saturation density \citep{2011PhRvC..83d5810D}. This means that one might expect the presence of a layer of hadronic perfect-fluid matter separating a quark phase from the elastic crust even in the case of sharp phase transitions. Thus, from Figure~\ref{tidal_def_1.4}, one would expect that relative tidal deformation changes are very small for systems with large perfect-fluid regions.

\begin{widetext}
\begin{center}
\begin{table}
[htbp] 
\caption{Hybrid NS properties for the models of Table \ref{ta_models}. Solid (elastic) phases start at $2\times 10^{14}$g cm$^{-3}$ or at the base of the hadronic phase (related to the relative tidal deformation changes of the last column) and end at $10^7$ g cm$^{-3}$ (for more explanation of the models, see the text). In our notation, $\Delta \Lambda \coloneqq \Lambda_{\mbox{perf}}- \Lambda_{\mbox{sol}}$. The radius of the quark phase is denoted by $R_{\mbox{quark}}$ whereas $\Delta R^{\mbox{sol}}_{\mbox{had}}\coloneqq R-r_{\epsilon=2\times 10^{14}}$ is the thickness of the elastic phase. For some $\eta$ and the HS1, HS2 and HS3 models, we have calculated relative tidal deformation changes for different masses (central densities, $\epsilon_c$) in order to check their dependencies on other quantities. For clarity, they are split by horizontal lines. All results for tidal deformations are related to $\tilde{\mu}$ as in Eq. \eqref{shear_modulus_model}, with $\tilde{\mu}_0=0$ and $\kappa=1.5\times 10^{-2}$. For the HS3 model, we have used the same parameters of Figure \ref{model1}.}
\begin{ruledtabular}
{\begin{tabular}{@{}ccccccccccccc@{}}
Hybrid & $\eta$ &  $\epsilon_c$ & $M$ & $R_{\mbox{quark}}$ & $R$ & $\Delta R_{\mbox{had}}^{\mbox{sol}}/R$ & $\Lambda_{\mbox{perf}}$&  $\Delta\Lambda/\Lambda_{\mbox{perf}}$ & $\left(\Delta\Lambda/\Lambda_{\mbox{perf}}\right)_{\mbox{had}}$ \\
model & & $(\epsilon_{\mbox{sat}})$ & (M$_{\odot})$ & (km) & (km) & \% & & & \\ \hline
HS1 & 0.0 & 1.530 & 0.43 & 1.06 & 11.71 & 38.1 & 2.558$\times 10^{5}$ & 2.32$\times 10^{-3}$ & 3.56$\times 10^{-2}$ \\
\textquotedbl & 0.16 & 1.485 & 0.37 & 1.19 & 11.90 & 45.0 & 5.984$\times 10^{5}$ & 5.25$\times 10^{-3}$ & 3.56$\times 10^{-2}$ \\
\textquotedbl & 0.30 & 1.46 & 0.33 & 1.02 & 11.86 & 51.5 & 1.148$\times 10^{6}$ & 9.92$\times 10^{-3}$& 3.56$\times 10^{-2}$  \\
\hline
\textquotedbl & 0.45 &1.9 & 0.95 & 9.28 & 11.84 & 16.6 & 4444.43 & 9.84$\times 10^{-5}$ & 3.59$\times 10^{-3}$  \\
 \textquotedbl & 0.45 &2.2 & 1.27 & 10.36 & 12.30 & 12.0 & 1041.92 & 4.29$\times 10^{-5}$ & 1.89$\times 10^{-3}$  \\
\textquotedbl  & 0.45 &2.358 & 1.40 & 10.69 & 12.43 & 10.7 & 629.5 & 3.25$\times 10^{-5}$ & 1.53$\times 10^{-3}$  \\
\textquotedbl  & 0.45 &3.0 & 1.73 & 11.29 & 12.62 & 8.0 & 178.61 & 1.65$\times 10^{-5}$ & 9.26$\times 10^{-4}$   \\
\textquotedbl  & 0.45 & 5.0 & 2.02 & 11.33 & 12.27 & 5.8 &42.24 & 7.57$\times 10^{-6}$ & 5.74$\times 10^{-4}$ \\
\hline
\textquotedbl & 0.60 & 1.442 & 0.27 & 2.25 & 11.90 & 64.6 & 3.314$\times 10^{6}$ & 2.34$\times 10^{-2}$ &3.47$\times 10^{-2}$ \\
\textquotedbl & 0.70 &\textquotedbl & 0.25 & 2.69 & 11.80 & 67.7 & 4.392$\times 10^{6}$ & 2.63$\times 10^{-2}$   &3.39$\times 10^{-2}$   \\
\textquotedbl & 0.80  & \textquotedbl &0.23 & 3.00 & 11.67 & 70.1 &5.695$\times 10^{6}$ & 2.83$\times 10^{-2}$ &3.33$\times 10^{-2}$    \\
\textquotedbl & 0.90 & \textquotedbl & 0.22 & 3.03 & 11.61 &73.9 & 7.223$\times 10^{6}$ & 3.07$\times 10^{-2}$& 3.56$\times 10^{-2}$ \\
\textquotedbl & 0.91 &\textquotedbl & 0.22 & 3.26 & 11.52 & 71.7 & 7.388$\times 10^{6}$ &  3.25$\times 10^{-2}$& 3.56$\times 10^{-2}$ \\
\textquotedbl & 1.0 & \textquotedbl & 0.20 & 3.43 & 11.36 &69.8 & 8.96$\times 10^{6}$ & 3.19$\times 10^{-2}$& 3.19$\times 10^{-2}$ \\
\textquotedbl & 1.5 & \textquotedbl & 0.16 & 3.99 & 11.47 & 61.9 & 1.924$\times 10^{7}$ &  2.81$\times 10^{-2}$ &  2.81$\times 10^{-2}$ \\
\hline
HS2 &  0.45 & 1.625 & 0.62 & 1.68 & 14.41 & 43.2 & 8.019$\times 10^{4}$ &  3.81$\times 10^{-3}$  & 3.57$\times 10^{-2}$ \\
\textquotedbl &  0.45 & 2.1 & 1.09 & 8.49 & 13.29 & 22.0 & 2593.44  & 3.30$\times 10^{-4}$ & 1.13$\times 10^{-2}$ \\
\textquotedbl &  0.45 & 2.483 & 1.40 & 9.7 & 13.24 & 16.3 & 643.21 & 1.30$\times 10^{-4}$ & 6.46$\times 10^{-3}$\\
\textquotedbl &  0.45 &3.1 & 1.69 & 10.43 & 13.13 & 12.5 & 196.28& 6.86$\times 10^{-5}$ & 4.03$\times 10^{-3}$ \\
\textquotedbl &  0.45 &5.1 & 1.98 & 10.68 & 12.55 & 9.0& 45.282 & 3.36$\times 10^{-5}$ & 2.42$\times 10^{-3}$ \\
\hline
HS3 & 0.77 & 3.925 & 1.12 & 2.71 & 12.32 & 14.1 & 1607.36 & 2.28$\times 10^{-5}$& 3.50$\times 10^{-2}$\\
\textquotedbl & 0.77 & 4.3 & 1.30 & 6.05 & 11.40 & 10.6 & 351.49 & 8.25$\times 10^{-5}$& 2.25$\times 10^{-2}$\\
\textquotedbl& 0.77 & 4.456 & 1.40 & 6.64 & 11.22 & 9.3 & 195.47 & 6.42$\times 10^{-6}$& 1.86$\times 10^{-2}$\\
\textquotedbl& 0.77 & 4.6& 1.49 & 7.03 & 11.09 & 8.4 & 123.99 & 5.88$\times 10^{-6}$& 1.60$\times 10^{-2}$\\
\textquotedbl& 0.77 & 5.0 & 1.68 & 7.71 & 10.88 & 6.7 & 48.033 & 1.20$\times 10^{-7}$& 1.14$\times 10^{-2}$ \\
\hline
\textquotedbl& 0.79 & 4.01 & 1.32 & 3.26 & 12.64 & 11.9 & 745.94 & 1.51$\times 10^{-5}$& 3.46$\times 10^{-2}$\\
\textquotedbl & 0.81 & 4.11 & 1.55 & 3.65 & 12.99 & 10.0 & 356.99 & 7.57$\times 10^{-6}$& 3.45$\times 10^{-2}$\\
\textquotedbl & 0.83 & 4.359 & 1.88 & 4.62 & 13.43 & 7.8 & 122.55 & 5.55$\times 10^{-6}$&3.37$\times 10^{-2}$\\
\textquotedbl & 0.83 & 5.2 & 2.01 & 7.09 & 12.25 & 6.0 & 32.229 & 3.18$\times 10^{-6}$& 2.42$\times 10^{-2}$ \\
\end{tabular} \label{ta1}}
\end{ruledtabular}
\\
\end{table}
\end{center}
\end{widetext}

Hybrid stars where an elastic hadronic phase would start right after the quark phase (and would extend until the ocean) could also be seen as a simplistic model for a solid star with an elastic mixed phase. Although we have not made any direct calculation in this regard, our numerical integrations with nonzero shear stresses right after the quark phase indicate that elastic mixed phases could lead to non-negligible tidal deformation changes (around $2-4\%$ or even larger). One would expect this, given that the shear modulus of a mixed phase might be much larger than the crustal shear modulus, still roughly $1-2\%$ of the local pressure \citep{2013NuPhA.906...37S}. However, there is an important, though subtle, point in this case. In the presence of a mixed phase, one would expect physical quantities such as the energy density and shear modulus to be continuous anywhere in the star. If one takes the quark phase as a perfect fluid, then $\tilde{\mu}$ should go continuously from zero to a nonzero value inside the mixed phase. This would render equations ill-defined at the quark-mixed phase interface since there are terms of the form $1/\tilde{\mu}$ in the perturbation equations. Clearly, the main source of the problem is the perfect-fluid approximation. If one takes the shear modulus to be small but finite, the tidal deformation equations for the mixed phase would be well defined. We leave this issue to be investigated elsewhere. In this case, boundary conditions for the problem change, which could have a non-negligible effect on tidal deformation changes when compared to perfect fluids. It would also be of interest to investigate the case of dynamical tides (see, e.g., \citealt{2019arXiv190608982A,2019arXiv190500012A,2019PhRvD.100b1501S} and references therein). Although they would change tidal deformations by some percent in the case of perfect fluids \citep{2019arXiv190608982A}, dynamical tides might have a larger impact on hybrid stars with solid phases due to aspects of the fundamental mode.

We have explicitly shown that surface degrees of freedom {cannot} be induced on phase-splitting surfaces. This is expected due to the assumption of no reactions around phase transitions and the static nature of the problem. It implies that there would be no way of significantly changing tidal deformations of solid crusts from the results we have found when the Einstein equations are taken into account. However, surface degrees of freedom could play an important role in more exotic cases with background degrees of freedom, such as boson stars or gravstars \citep{2018arXiv180408026J}.

Our results suggest that when the accuracy of tidal deformations is around 5$\%$, even in the most conservative cases, shear stresses should not be disregarded. In the worst case, they would be an important part of systematic uncertainties for some models. It would be of interest to advance the analysis proposed here for more realistic hybrid EOS in order to single out particularities lost in our idealized approach. Most importantly, one should also calculate consistently the shear modulus for a given EOS.

\section{Acknowledgements}
We acknowledge the constructive discussions with Leszek Zdunik, Pawe{\l} Haensel, Morgane Fortin and Lami Souleiman. This work was partially supported by the Polish National Science Centre grant no. 2016/22/E/ST9/00037. JPP is also thankful for the partial support given by FAPESP (grants number 2015/04174-9 and 2017/21384-2). NA acknowledges financial support from STFC via grant no. ST/R00045X/1.

\bibliographystyle{apj}
\bibliography{tidal_deformations}

\begin{thebibliography}{90}
\expandafter\ifx\csname natexlab\endcsname\relax\def\natexlab#1{#1}\fi

\bibitem[{{Abbott} {et~al.}(2016{\natexlab{a}}){Abbott}, {Abbott}, {Abbott},
  {Abernathy}, {Acernese}, {Ackley}, {Adams}, {Adams}, {Addesso}, {Adhikari},
  {et~al.}}]{2016PhRvL.116x1103A}
{Abbott}, B.~P., {Abbott}, R., {Abbott}, T.~D., {et~al.} 2016{\natexlab{a}},
  Phys. Rev. Lett., 116, 241103

\bibitem[{{Abbott} {et~al.}(2016{\natexlab{b}}){Abbott}, {Abbott}, {Abbott},
  {Abernathy}, {Acernese}, {Ackley}, {Adams}, {Adams}, {Addesso}, {Adhikari},
  {et~al.}}]{2016PhRvL.116f1102A}
---. 2016{\natexlab{b}}, Phys. Rev. Lett., 116, 061102

\bibitem[{{Abbott} {et~al.}(2017{\natexlab{a}}){Abbott}, {Abbott}, {Abbott},
  {Abernathy}, {Ackley}, {Adams}, {Addesso}, {Adhikari},
  {et~al.}}]{2017CQGra..34d4001A}
---. 2017{\natexlab{a}}, Classical and Quantum Gravity, 34, 044001

\bibitem[{{Abbott} {et~al.}(2017{\natexlab{b}}){Abbott}, {Abbott}, {Abbott},
  {Abernathy}, {Acernese}, {Ackley}, {Adams}, {Adams}, {Addesso}, {Adhikari},
  {et~al.}}]{PhysRevLett.118.221101}
---. 2017{\natexlab{b}}, Phys. Rev. Lett., 118, 221101

\bibitem[{{Abbott} {et~al.}(2017{\natexlab{c}}){Abbott}, {Abbott}, {Abbott},
  {Acernese}, {Ackley}, {Adams}, {Adams}, {Addesso}, {Adhikari},
  {et~al.}}]{2017PhRvL.119p1101A}
---. 2017{\natexlab{c}}, Physical Review Letters, 119, 161101

\bibitem[{{Abbott} {et~al.}(2018){Abbott}, {Abbott}, {Abbott}, {Acernese},
  {Ackley}, {Adams}, {Adams}, {Addesso}, {et~al.}}]{2018arXiv180511581T}
---. 2018, \prl, 121, 161101

\bibitem[{{Abbott} {et~al.}(2019{\natexlab{a}}){Abbott}, {Abbott}, {Abbott},
  {Abernathy}, {Acernese}, {Ackley}, {Adams}, {Adams}, {Addesso}, {Adhikari},
  {et~al.}}]{2019PhRvX...9c1040A}
---. 2019{\natexlab{a}}, Physical Review X, 9, 031040

\bibitem[{{Abbott} {et~al.}(2019{\natexlab{b}}){Abbott}, {Abbott}, {Abbott},
  {Acernese}, {Ackley}, {Adams}, {Adams}, {Addesso}, {Adhikari},
  {et~al.}}]{2019PhRvX...9a1001A}
---. 2019{\natexlab{b}}, Physical Review X, 9, 011001

\bibitem[{{Abbott} {et~al.}(2020)}]{2020arXiv200101761T}
{Abbott}, B.~P., {et~al.} 2020, arXiv e-prints, arXiv:2001.01761

\bibitem[{{Alford} {et~al.}(2005){Alford}, {Braby}, {Paris}, \&
  {Reddy}}]{2005ApJ...629..969A}
{Alford}, M., {Braby}, M., {Paris}, M., \& {Reddy}, S. 2005, \apj, 629, 969

\bibitem[{{Alford} \& {Sedrakian}(2017)}]{2017PhRvL.119p1104A}
{Alford}, M., \& {Sedrakian}, A. 2017, Physical Review Letters, 119, 161104

\bibitem[{Alford {et~al.}(2019)Alford, Han, \& Schwenzer}]{Alford:2019oge}
Alford, M.~G., Han, S., \& Schwenzer, K. 2019, J. Phys., G46, 114001

\bibitem[{{Alford} {et~al.}(2008){Alford}, {Schmitt}, {Rajagopal}, \&
  {Sch{\"a}fer}}]{2008RvMP...80.1455A}
{Alford}, M.~G., {Schmitt}, A., {Rajagopal}, K., \& {Sch{\"a}fer}, T. 2008,
  Reviews of Modern Physics, 80, 1455

\bibitem[{{Alvarez-Castillo} \& {Blaschke}(2017)}]{2017PhRvC..96d5809A}
{Alvarez-Castillo}, D.~E., \& {Blaschke}, D.~B. 2017, \prc, 96, 045809

\bibitem[{{Andersson} \& {Comer}(2007)}]{2007LRR....10....1A}
{Andersson}, N., \& {Comer}, G.~L. 2007, Living Reviews in Relativity, 10, 1

\bibitem[{{Andersson} {et~al.}(2019){Andersson}, {Haskell}, {Comer}, \&
  {Samuelsson}}]{2019CQGra..36j5004A}
{Andersson}, N., {Haskell}, B., {Comer}, G.~L., \& {Samuelsson}, L. 2019,
  Classical and Quantum Gravity, 36, 105004

\bibitem[{{Andersson} \& {Pnigouras}(2019{\natexlab{a}})}]{2019arXiv190500012A}
{Andersson}, N., \& {Pnigouras}, P. 2019{\natexlab{a}}, arXiv e-prints,
  arXiv:1905.00012

\bibitem[{{Andersson} \& {Pnigouras}(2019{\natexlab{b}})}]{2019arXiv190608982A}
---. 2019{\natexlab{b}}, arXiv e-prints, arXiv:1906.08982

\bibitem[{{Annala} {et~al.}(2018){Annala}, {Gorda}, {Kurkela}, \&
  {Vuorinen}}]{2018PhRvL.120q2703A}
{Annala}, E., {Gorda}, T., {Kurkela}, A., \& {Vuorinen}, A. 2018, Physical
  Review Letters, 120, 172703

\bibitem[{{Baiotti}(2019)}]{2019PrPNP.10903714B}
{Baiotti}, L. 2019, Progress in Particle and Nuclear Physics, 109, 103714

\bibitem[{{Bauswein} {et~al.}(2018){Bauswein}, {Bastian}, {Blaschke},
  {Chatziioannou}, {Clark}, {Fischer}, \& {Oertel}}]{2018arXiv180901116B}
{Bauswein}, A., {Bastian}, N.-U.~F., {Blaschke}, D.~B., {et~al.} 2018, ArXiv
  e-prints

\bibitem[{{Bauswein} {et~al.}(2019){Bauswein}, {Bastian}, {Blaschke},
  {Chatziioannou}, {Clark}, {Fischer}, \& {Oertel}}]{2019PhRvL.122f1102B}
---. 2019, \prl, 122, 061102

\bibitem[{{Baym} {et~al.}(1971){Baym}, {Pethick}, \&
  {Sutherland}}]{1971ApJ...170..299B}
{Baym}, G., {Pethick}, C., \& {Sutherland}, P. 1971, \apj, 170, 299

\bibitem[{{Bejger} {et~al.}(2017){Bejger}, {Blaschke}, {Haensel}, {Zdunik}, \&
  {Fortin}}]{2017A&A...600A..39B}
{Bejger}, M., {Blaschke}, D., {Haensel}, P., {Zdunik}, J.~L., \& {Fortin}, M.
  2017, \aap, 600, A39

\bibitem[{{Bilous} {et~al.}(2019){Bilous}, {Watts}, {Harding}, {Riley},
  {Arzoumanian}, {Bogdanov}, {Gendreau}, {Ray}, {Guillot}, {Ho}, \&
  {Chakrabarty}}]{Bilous_2019}
{Bilous}, A.~V., {Watts}, A.~L., {Harding}, A.~K., {et~al.} 2019, \apjl, 887,
  L23

\bibitem[{{Bogdanov} {et~al.}(2019{\natexlab{a}}){Bogdanov}, {Guillot}, {Ray},
  {Wolff}, {Chakrabarty}, {Ho}, {Kerr}, {Lamb}, {Lommen}, {Ludlam}, {Milburn},
  {Montano}, {Miller}, {Baub{\"o}ck}, {{\"O}zel}, {Psaltis}, {Remillard},
  {Riley}, {Steiner}, {Strohmayer}, {Watts}, {Wood}, {Zeldes}, {Enoto},
  {Okajima}, {Kellogg}, {Baker}, {Markwardt}, {Arzoumanian}, \&
  {Gendreau}}]{Bogdanov_2019}
{Bogdanov}, S., {Guillot}, S., {Ray}, P.~S., {et~al.} 2019{\natexlab{a}},
  \apjl, 887, L25

\bibitem[{{Bogdanov} {et~al.}(2019{\natexlab{b}}){Bogdanov}, {Lamb},
  {Mahmoodifar}, {Miller}, {Morsink}, {Riley}, {Strohmayer}, {Tung}, {Watts},
  {Dittmann}, {Chakrabarty}, {Guillot}, {Arzoumanian}, \&
  {Gendreau}}]{Bogdanov_2019_}
{Bogdanov}, S., {Lamb}, F.~K., {Mahmoodifar}, S., {et~al.} 2019{\natexlab{b}},
  \apjl, 887, L26

\bibitem[{Bombaci {et~al.}(2016)Bombaci, Logoteta, Vida\~na, \&
  Provid\^encia}]{Bombaci:2016xuj}
Bombaci, I., Logoteta, D., Vida\~na, I., \& Provid\^encia, C. 2016, Eur. Phys.
  J. A, 52, 58

\bibitem[{{Chamel} \& {Haensel}(2008)}]{2008LRR....11...10C}
{Chamel}, N., \& {Haensel}, P. 2008, Living Reviews in Relativity, 11, 10

\bibitem[{Chatziioannou {et~al.}(2017)Chatziioannou, Klein, Yunes, \&
  Cornish}]{PhysRevD.95.104004}
Chatziioannou, K., Klein, A., Yunes, N., \& Cornish, N. 2017, Phys. Rev. D, 95,
  104004

\bibitem[{{Chirenti} {et~al.}(2017){Chirenti}, {Gold}, \&
  {Miller}}]{2017ApJ...837...67C}
{Chirenti}, C., {Gold}, R., \& {Miller}, M.~C. 2017, \apj, 837, 67

\bibitem[{{Christian} \& {Schaffner-Bielich}(2019)}]{2019arXiv191209809C}
{Christian}, J.-E., \& {Schaffner-Bielich}, J. 2019, arXiv e-prints,
  arXiv:1912.09809

\bibitem[{{Coughlin} {et~al.}(2019){Coughlin}, {Dietrich}, {Margalit}, \&
  {Metzger}}]{2019MNRAS.489L..91C}
{Coughlin}, M.~W., {Dietrich}, T., {Margalit}, B., \& {Metzger}, B.~D. 2019,
  \mnras, 489, L91

\bibitem[{{Cutler} {et~al.}(2003){Cutler}, {Ushomirsky}, \&
  {Link}}]{2003ApJ...588..975C}
{Cutler}, C., {Ushomirsky}, G., \& {Link}, B. 2003, \apj, 588, 975

\bibitem[{{Damour} \& {Nagar}(2009)}]{2009PhRvD..80h4035D}
{Damour}, T., \& {Nagar}, A. 2009, \prd, 80, 084035

\bibitem[{De {et~al.}(2018)De, Finstad, Lattimer, Brown, Berger, \&
  Biwer}]{PhysRevLett.121.091102}
De, S., Finstad, D., Lattimer, J.~M., {et~al.} 2018, Phys. Rev. Lett., 121,
  091102

\bibitem[{{de Lima} {et~al.}(2020){de Lima}, {Coelho}, {Pereira}, {Rodrigues},
  \& {Rueda}}]{2020ApJ...889..165D}
{de Lima}, R. C.~R., {Coelho}, J.~G., {Pereira}, J.~P., {Rodrigues}, C.~V., \&
  {Rueda}, J.~A. 2020, \apj, 889, 165

\bibitem[{{Douchin} \& {Haensel}(2001)}]{2001A&A...380..151D}
{Douchin}, F., \& {Haensel}, P. 2001, \aap, 380, 151

\bibitem[{{Ducoin} {et~al.}(2011){Ducoin}, {Margueron}, {Provid{\^e}ncia}, \&
  {Vida{\~n}a}}]{2011PhRvC..83d5810D}
{Ducoin}, C., {Margueron}, J., {Provid{\^e}ncia}, C., \& {Vida{\~n}a}, I. 2011,
  \prc, 83, 045810

\bibitem[{{Essick} {et~al.}(2019){Essick}, {Landry}, \&
  {Holz}}]{2019arXiv191009740E}
{Essick}, R., {Landry}, P., \& {Holz}, D.~E. 2019, arXiv e-prints,
  arXiv:1910.09740

\bibitem[{{Finn}(1990)}]{1990MNRAS.245...82F}
{Finn}, L.~S. 1990, \mnras, 245, 82

\bibitem[{{Gittins} {et~al.}(2020){Gittins}, {Andersson}, \&
  {Pereira}}]{2020arXiv200305449G}
{Gittins}, F., {Andersson}, N., \& {Pereira}, J.~P. 2020, arXiv e-prints,
  arXiv:2003.05449

\bibitem[{{Guillot} {et~al.}(2019){Guillot}, {Kerr}, {Ray}, {Bogdanov},
  {Ransom}, {Deneva}, {Arzoumanian}, {Bult}, {Chakrabarty}, {Gendreau}, {Ho},
  {Jaisawal}, {Malacaria}, {Miller}, {Strohmayer}, {Wolff}, {Wood}, {Webb},
  {Guillemot}, {Cognard}, \& {Theureau}}]{Guillot_2019}
{Guillot}, S., {Kerr}, M., {Ray}, P.~S., {et~al.} 2019, \apjl, 887, L27

\bibitem[{{Haensel} {et~al.}(2007){Haensel}, {Potekhin}, \&
  {Yakovlev}}]{2007ASSL..326.....H}
{Haensel}, P., {Potekhin}, A.~Y., \& {Yakovlev}, D.~G. 2007, {Neutron Stars 1 :
  Equation of State and Structure}, Vol. 326 (Springer)

\bibitem[{{Han} \& {Steiner}(2019)}]{2019PhRvD..99h3014H}
{Han}, S., \& {Steiner}, A.~W. 2019, \prd, 99, 083014

\bibitem[{{Haskell} {et~al.}(2007){Haskell}, {Andersson}, {Jones}, \&
  {Samuelsson}}]{2007PhRvL..99w1101H}
{Haskell}, B., {Andersson}, N., {Jones}, D.~I., \& {Samuelsson}, L. 2007,
  Physical Review Letters, 99, 231101

\bibitem[{{Haskell} {et~al.}(2006){Haskell}, {Jones}, \&
  {Andersson}}]{2006MNRAS.373.1423H}
{Haskell}, B., {Jones}, D.~I., \& {Andersson}, N. 2006, \mnras, 373, 1423

\bibitem[{{Hinderer}(2008)}]{2008ApJ...677.1216H}
{Hinderer}, T. 2008, \apj, 677, 1216

\bibitem[{{Hinderer} {et~al.}(2010){Hinderer}, {Lackey}, {Lang}, \&
  {Read}}]{2010PhRvD..81l3016H}
{Hinderer}, T., {Lackey}, B.~D., {Lang}, R.~N., \& {Read}, J.~S. 2010, \prd,
  81, 123016

\bibitem[{{Jie Li} {et~al.}(2019){Jie Li}, {Sedrakian}, \&
  {Alford}}]{2019arXiv191100276J}
{Jie Li}, J., {Sedrakian}, A., \& {Alford}, M. 2019, arXiv e-prints,
  arXiv:1911.00276

\bibitem[{Jim\'enez~Forteza {et~al.}(2018)Jim\'enez~Forteza, Abdelsalhin, Pani,
  \& Gualtieri}]{PhysRevD.98.124014}
Jim\'enez~Forteza, X., Abdelsalhin, T., Pani, P., \& Gualtieri, L. 2018, Phys.
  Rev. D, 98, 124014

\bibitem[{{Johnson-McDaniel} {et~al.}(2018){Johnson-McDaniel}, {Mukherjee},
  {Kashyap}, {Ajith}, {Del Pozzo}, \& {Vitale}}]{2018arXiv180408026J}
{Johnson-McDaniel}, N.~K., {Mukherjee}, A., {Kashyap}, R., {et~al.} 2018, arXiv
  e-prints, arXiv:1804.08026

\bibitem[{{Kiuchi} {et~al.}(2019){Kiuchi}, {Kyutoku}, {Shibata}, \&
  {Taniguchi}}]{2019ApJ...876L..31K}
{Kiuchi}, K., {Kyutoku}, K., {Shibata}, M., \& {Taniguchi}, K. 2019, \apjl,
  876, L31

\bibitem[{{Kr{\"u}ger} {et~al.}(2015){Kr{\"u}ger}, {Ho}, \&
  {Andersson}}]{2015PhRvD..92f3009K}
{Kr{\"u}ger}, C.~J., {Ho}, W.~C.~G., \& {Andersson}, N. 2015, \prd, 92, 063009

\bibitem[{{Landau} \& {Lifshitz}(1975)}]{1975ctf..book.....L}
{Landau}, L.~D., \& {Lifshitz}, E.~M. 1975 (Pergamon Press, Oxford)

\bibitem[{{Lau} {et~al.}(2017){Lau}, {Leung}, \& {Lin}}]{2017PhRvD..95j1302L}
{Lau}, S.~Y., {Leung}, P.~T., \& {Lin}, L.-M. 2017, \prd, 95, 101302

\bibitem[{{Lau} {et~al.}(2019){Lau}, {Leung}, \& {Lin}}]{2019PhRvD..99b3018L}
{Lau}, S.~Y., {Leung}, P.~T., \& {Lin}, L.~M. 2019, \prd, 99, 023018

\bibitem[{Lugones \& Grunfeld(2017)}]{Lugones:2016ytl}
Lugones, G., \& Grunfeld, A.~G. 2017, Phys. Rev., C95, 015804

\bibitem[{{Maggiore} {et~al.}(2019){Maggiore}, {van den Broeck}, {Bartolo},
  {Belgacem}, {Bertacca}, {Bizouard}, {Branchesi}, {Clesse}, {Foffa},
  {Garc{\'\i}a-Bellido}, {Grimm}, {Harms}, {Hinderer}, {Matarrese}, {Palomba},
  {Peloso}, {Ricciardone}, \& {Sakellariadou}}]{2019arXiv191202622M}
{Maggiore}, M., {van den Broeck}, C., {Bartolo}, N., {et~al.} 2019, arXiv
  e-prints, arXiv:1912.02622

\bibitem[{{Mannarelli} {et~al.}(2007){Mannarelli}, {Rajagopal}, \&
  {Sharma}}]{2007PhRvD..76g4026M}
{Mannarelli}, M., {Rajagopal}, K., \& {Sharma}, R. 2007, \prd, 76, 074026

\bibitem[{{Miller} {et~al.}(2019){Miller}, {Lamb}, {Dittmann}, {Bogdanov},
  {Arzoumanian}, {Gendreau}, {Guillot}, {Harding}, {Ho}, {Lattimer}, {Ludlam},
  {Mahmoodifar}, {Morsink}, {Ray}, {Strohmayer}, {Wood}, {Enoto}, {Foster},
  {Okajima}, {Prigozhin}, \& {Soong}}]{Miller_2019}
{Miller}, M.~C., {Lamb}, F.~K., {Dittmann}, A.~J., {et~al.} 2019, \apjl, 887,
  L24

\bibitem[{{Monta{\~n}a} {et~al.}(2019){Monta{\~n}a}, {Tol{\'o}s}, {Hanauske},
  \& {Rezzolla}}]{2019PhRvD..99j3009M}
{Monta{\~n}a}, G., {Tol{\'o}s}, L., {Hanauske}, M., \& {Rezzolla}, L. 2019,
  \prd, 99, 103009

\bibitem[{{Most} {et~al.}(2018){Most}, {Weih}, {Rezzolla}, \&
  {Schaffner-Bielich}}]{2018PhRvL.120z1103M}
{Most}, E.~R., {Weih}, L.~R., {Rezzolla}, L., \& {Schaffner-Bielich}, J. 2018,
  Physical Review Letters, 120, 261103

\bibitem[{{Nandi} \& {Char}(2018)}]{2018ApJ...857...12N}
{Nandi}, R., \& {Char}, P. 2018, \apj, 857, 12

\bibitem[{{{\"O}zel} {et~al.}(2016){{\"O}zel}, {Psaltis}, {Arzoumanian},
  {Morsink}, \& {Baub{\"o}ck}}]{2016ApJ...832...92O}
{{\"O}zel}, F., {Psaltis}, D., {Arzoumanian}, Z., {Morsink}, S., \&
  {Baub{\"o}ck}, M. 2016, \apj, 832, 92

\bibitem[{Paschalidis {et~al.}(2018)Paschalidis, Yagi, Alvarez-Castillo,
  Blaschke, \& Sedrakian}]{Paschalidis2018}
Paschalidis, V., Yagi, K., Alvarez-Castillo, D., Blaschke, D.~B., \& Sedrakian,
  A. 2018, Phys. Rev. D, 97, 084038

\bibitem[{{Penner} {et~al.}(2011){Penner}, {Andersson}, {Samuelsson}, {Hawke},
  \& {Jones}}]{2011PhRvD..84j3006P}
{Penner}, A.~J., {Andersson}, N., {Samuelsson}, L., {Hawke}, I., \& {Jones},
  D.~I. 2011, \prd, 84, 103006

\bibitem[{{Pereira} {et~al.}(2014){Pereira}, {Coelho}, \&
  {Rueda}}]{2014PhRvD..90l3011P}
{Pereira}, J.~P., {Coelho}, J.~G., \& {Rueda}, J.~A. 2014, \prd, 90, 123011

\bibitem[{{Pereira} {et~al.}(2018){Pereira}, {Flores}, \&
  {Lugones}}]{2018ApJ...860...12P}
{Pereira}, J.~P., {Flores}, C.~V., \& {Lugones}, G. 2018, \apj, 860, 12

\bibitem[{{Pereira} \& {Lugones}(2019)}]{2019ApJ...871...47P}
{Pereira}, J.~P., \& {Lugones}, G. 2019, \apj, 871, 47

\bibitem[{{Poisson}(2004)}]{2004reto.book.....P}
{Poisson}, E. 2004 (Cambridge University Press, Cambridge)

\bibitem[{{Postnikov} {et~al.}(2010){Postnikov}, {Prakash}, \&
  {Lattimer}}]{2010PhRvD..82b4016P}
{Postnikov}, S., {Prakash}, M., \& {Lattimer}, J.~M. 2010, \prd, 82, 024016

\bibitem[{{Punturo} {et~al.}(2010){Punturo}, {Abernathy}, {Acernese}, {Allen},
  {Andersson}, {Arun}, {Barone}, {Barr}, {Barsuglia}, {Beker}, {Beveridge},
  {Birindelli}, {Bose}, {Bosi}, {Braccini}, {Bradaschia}, {Bulik}, {Calloni},
  {Cella}, {et~al.}}]{2010CQGra..27s4002P}
{Punturo}, M., {Abernathy}, M., {Acernese}, F., {et~al.} 2010, Classical and
  Quantum Gravity, 27, 194002

\bibitem[{{Raaijmakers} {et~al.}(2019){Raaijmakers}, {Riley}, {Watts}, {Greif},
  {Morsink}, {Hebeler}, {Schwenk}, {Hinderer}, {Nissanke}, {Guillot},
  {Arzoumanian}, {Bogdanov}, {Chakrabarty}, {Gendreau}, {Ho}, {Lattimer},
  {Ludlam}, \& {Wolff}}]{Raaijmakers_2019}
{Raaijmakers}, G., {Riley}, T.~E., {Watts}, A.~L., {et~al.} 2019, \apjl, 887,
  L22

\bibitem[{{Radice} \& {Dai}(2019)}]{2019EPJA...55...50R}
{Radice}, D., \& {Dai}, L. 2019, European Physical Journal A, 55, 50

\bibitem[{{Radice} {et~al.}(2018){Radice}, {Perego}, {Zappa}, \&
  {Bernuzzi}}]{2018ApJ...852L..29R}
{Radice}, D., {Perego}, A., {Zappa}, F., \& {Bernuzzi}, S. 2018, \apjl, 852,
  L29

\bibitem[{{Regge} \& {Wheeler}(1957)}]{1957PhRv..108.1063R}
{Regge}, T., \& {Wheeler}, J.~A. 1957, Physical Review, 108, 1063

\bibitem[{{Riley} {et~al.}(2019){Riley}, {Watts}, {Bogdanov}, {Ray}, {Ludlam},
  {Guillot}, {Arzoumanian}, {Baker}, {Bilous}, {Chakrabarty}, {Gendreau},
  {Harding}, {Ho}, {Lattimer}, {Morsink}, \& {Strohmayer}}]{Riley_2019}
{Riley}, T.~E., {Watts}, A.~L., {Bogdanov}, S., {et~al.} 2019, \apjl, 887, L21

\bibitem[{{Schmidt} \& {Hinderer}(2019)}]{2019PhRvD.100b1501S}
{Schmidt}, P., \& {Hinderer}, T. 2019, \prd, 100, 021501

\bibitem[{{Shapiro} \& {Teukolsky}(1986)}]{1986bhwd.book.....S}
{Shapiro}, S.~L., \& {Teukolsky}, S.~A. 1986, {Black Holes, White Dwarfs and
  Neutron Stars: The Physics of Compact Objects} (Willey-CH)

\bibitem[{{Sieniawska} {et~al.}(2018){Sieniawska}, {Bejger}, \&
  {Haskell}}]{2018A&A...616A.105S}
{Sieniawska}, M., {Bejger}, M., \& {Haskell}, B. 2018, \aap, 616, A105

\bibitem[{{Sieniawska} {et~al.}(2019){Sieniawska}, {Turcza{\'n}ski}, {Bejger},
  \& {Zdunik}}]{2019A&A...622A.174S}
{Sieniawska}, M., {Turcza{\'n}ski}, W., {Bejger}, M., \& {Zdunik}, J.~L. 2019,
  \aap, 622, A174

\bibitem[{{Sotani} {et~al.}(2007){Sotani}, {Kokkotas}, \&
  {Stergioulas}}]{2007MNRAS.375..261S}
{Sotani}, H., {Kokkotas}, K.~D., \& {Stergioulas}, N. 2007, \mnras, 375, 261

\bibitem[{{Sotani} {et~al.}(2013){Sotani}, {Maruyama}, \&
  {Tatsumi}}]{2013NuPhA.906...37S}
{Sotani}, H., {Maruyama}, T., \& {Tatsumi}, T. 2013, \nphysa, 906, 37

\bibitem[{{Tooper}(1965)}]{Tooper1965}
{Tooper}, R.~F. 1965, \apj, 142, 1541

\bibitem[{{Ushomirsky} {et~al.}(2000){Ushomirsky}, {Cutler}, \&
  {Bildsten}}]{2000MNRAS.319..902U}
{Ushomirsky}, G., {Cutler}, C., \& {Bildsten}, L. 2000, \mnras, 319, 902

\bibitem[{Vallisneri(2008)}]{PhysRevD.77.042001}
Vallisneri, M. 2008, Phys. Rev. D, 77, 042001

\bibitem[{{Zdunik}(2000)}]{Zdunik2000}
{Zdunik}, J.~L. 2000, \aap, 359, 311

\bibitem[{{Zdunik} {et~al.}(2008){Zdunik}, {Bejger}, \&
  {Haensel}}]{2008A&A...491..489Z}
{Zdunik}, J.~L., {Bejger}, M., \& {Haensel}, P. 2008, \aap, 491, 489

\bibitem[{{Zhang} {et~al.}(2016){Zhang}, {Feroci},
  {et~al.}}]{2016SPIE.9905E..1QZ}
{Zhang}, S.~N., {Feroci}, M., {et~al.} 2016, in \procspie, Vol. 9905, Space
  Telescopes and Instrumentation 2016: Ultraviolet to Gamma Ray, 99051Q

\end{thebibliography}

\end{document}